\documentclass[amsfonts]{elsarticle}

\usepackage{graphicx}
\usepackage{amsmath,amsfonts}
\usepackage{epsfig,color}

\begin{document}
\newcommand{\msb}{\overline{MS}}
\newcommand{\be}{\begin{eqnarray}}
\newcommand{\ee}{\end{eqnarray}}
\newcommand\del{\partial}
\newcommand\nn{\nonumber}
\newcommand{\Tr}{{\rm Tr}}
\newcommand{\Str}{{\rm Str\,}}
\newcommand{\Sdet}{{\rm Sdet\,}}
 \newcommand{\Pf}{{\rm Pf\,}}
\newcommand{\U}{{\rm U\,}}
\newcommand{\mat}{\left ( \begin{array}{cc}}
\newcommand{\emat}{\end{array} \right )}
\newcommand{\vect}{\left ( \begin{array}{c}}
\newcommand{\evect}{\end{array} \right )}
\newcommand{\tr}{{\rm Tr}}
\newcommand{\hm}{\hat m}
\newcommand{\ha}{\hat a}
\newcommand{\hz}{\hat z}
\newcommand{\hx}{\hat x}
\newcommand{\hl}{\hat{\lambda}}
\newcommand{\tm}{\tilde{m}}
\newcommand{\ta}{\tilde{a}}
\newcommand{\tz}{\tilde{z}}
\newcommand{\tx}{\tilde{x}}
\definecolor{red}{rgb}{1.00, 0.00, 0.00}
\newcommand{\rd}{\color{red}}
\definecolor{blue}{rgb}{0.00, 0.00, 1.00}
\definecolor{green}{rgb}{0.10, 1.00, .10}
\newcommand{\blu}{\color{blue}}
\newcommand{\green}{\color{green}}



\title{Wilson chiral perturbation theory for dynamical twisted mass fermions vs lattice data - a case study }
\author{Krzysztof Cichy}
\address{Faculty of Physics, Adam Mickiewicz University, Umultowska
85, 61-614 Pozna\'{n}, Poland}
  
\author{Savvas Zafeiropoulos}
\address{Institute for Theoretical Physics, Heidelberg University, Philosophenweg 12, 69120 Heidelberg,
Germany}

\begin  {abstract}
We compute the low lying eigenvalues of the Hermitian Dirac operator in lattice QCD with $N_{\rm f} = 2+1+1$ twisted mass fermions.
We discuss whether these eigenvalues are in the $\epsilon$-regime or the $p$-regime of Wilson chiral perturbation theory ($\chi$PT) for twisted mass fermions.
Reaching the deep $\epsilon$-regime is practically unfeasible with presently typical simulation parameters, but still the few lowest eigenvalues of the employed ensemble evince some characteristic $\epsilon$-regime features.
With this conclusion in mind, we develop a fitting strategy to extract two low energy constants from analytical $\epsilon$-regime predictions at a fixed index.
Thus, we obtain results for the chiral condensate and the low energy constant $W_8$. 
We also discuss how to improve both the theoretical calculation and the lattice computation.
\end{abstract}
\date{\today}

\maketitle

\section{Introduction}

The systematic control and elimination of discretization errors has been the center of attention for the community of lattice field theories in the past decades. The twisted mass formulation of lattice QCD is one of the most successful ways to improve the cutoff effects of the Wilson discretization~\cite{FrezzottiA, Frezzossi, Frezzossi2}. At maximal twist, the discretization errors in the action and the matrix elements are of $\mathcal{O}(a^2)$. The other main advantage of the twisted mass prescription is the absence of exceptional configurations, since the problem of small eigenvalues of the Hermitian Dirac operator is regulated by the addition of the twisted mass. This is therefore a very promising description which allows for fast simulations of dynamical fermions with solid theoretical foundations. The main drawback of this prescription is that parity and isospin symmetry are broken by cutoff effects of $\mathcal{O}(a^2)$. For a pedagogical and detailed introduction to the twisted mass formulation, we refer the reader to ~\cite{Sint, Andrea}. 

 Recently, the European Twisted Mass collaboration (ETMC) has been simulating twisted mass fermions at the physical pion mass~\cite{PP1, PP2}, which already gets rid of one extrapolation, namely the chiral one. However, physical point simulations come at a heavy price and simulations at a heavier pion mass are still performed. During the course of a lattice study, there is another necessary extrapolation to be made and it is the one to the continuum limit. When simulating closer and closer to the continuum limit, one faces extremely severe problems, mainly related to critical slowing down~\cite{slowing} and the freezing of topology. The introduction of open boundary conditions~\cite{LSopen} significantly ameliorated the issue, but still simulations with values of the lattice spacing $<0.05$ fm remain very difficult. So one needs to perform a combined chiral and continuum extrapolation and we advocate here for a lattice augmented version of the low energy Effective Field Theory (EFT) for QCD, which correctly incorporates discretization errors to leading order (LO) in $a$. 
 
 The low-energy EFT for Wilson fermions, Wilson $\chi$PT, was introduced in ~\cite{SS,RS, BRS}. It provides a systematic framework to study the quark mass dependence as well as the discretization effects in various, phenomenologically interesting observables. We refer the reader to~\cite{SharpeNara} for a detailed and pedagogical introduction to Wilson $\chi$PT. Moreover, one can study the intricate phase diagram of twisted mass fermions in analytical mean field studies employing Wilson $\chi$PT~\cite{SharpeWu, Scorzato, Muenster, KSVZ, JKSVZ}. When $m\propto a^2$, one has two possibilities, that of a second order phase transition to the so called Aoki phase~\cite{Aoki} or a first order scenario (the Sharpe-Singleton scenario) \cite{SS}.  The sign and the strength of the new low energy constants (LECs) which parametrize lattice artifacts determine which of the two scenarios is realized in practice during a lattice simulation, see e.g.\ \cite{realization}. By the same token, one can describe within Wilson $\chi$PT the changes in the orientation of the chiral condensate \cite{SharpeWu, SharpeNara, JKSVZ}. The chiral condensate changes promptly from -1 to +1 in the first order scenario, while it changes in a continuous manner in the Aoki phase. Finally, the extraction of the physical LECs, such as $F_{\pi}$ and $\Sigma$, through fits to $\chi$PT formulae hinges strongly on the knowledge of the LECs of Wilson  $\chi$PT.
 
  Consequently, there has been a great deal of analytical~\cite{DSV, ADSV, KVZprl, Hansen, KVZprd, KVZ2c, AokiBar, B1, B2, B3} and numerical~\cite{Krzmixed, splittings, Fabio, DHS1, DWW, DHS2} work on the extraction of the LECs of Wilson $\chi$PT. The LECs of Wilson $\chi$PT parametrize the pion mass splittings~\cite{splittings}, the difference of the pion scattering lengths between channels with isospin zero and isospin equal to two~\cite{AokiBar, Fabio} and they also measure the departure from unitarity in a mixed action setup where one simulates overlap fermions in a sea of a cheaper discretization, e.g.\ twisted mass fermions~\cite{Krzmixed}. 

A very promising method, which we will follow in this paper, is to extract the LECs of Wilson $\chi$PT by fitting analytical results derived in the framework of Wilson $\chi$PT in the $\epsilon$-regime to eigenvalue densities of the Dirac operator computed on the lattice. We compute numerically the microscopic spectral density for lattice QCD with twisted mass fermions and compare it with the analytical result  presented in \cite{TMSV}. We obtain results for the chiral condensate and the low-energy constant $W_8$ of Wilson $\chi$PT by fitting the lattice data for the microscopic spectral density of the Hermitian Wilson Dirac operator with a fixed index and at a finite volume to the analytical results. This is a case study at one value of the lattice spacing and at one volume, attempting, for the first time for dynamical twisted mass fermions, to study numerically the spectrum of the twisted mass Wilson Dirac operator, test the validity of the analytic results of Wilson $\chi$PT and extract directly from the spectrum two important low-energy constants. Preliminary results were presented in \cite{lat15proc,proc2}. 
Note that the microscopic eigenvalue density is extremely sensitive to 
the ${\cal O}(a^2)$ effects of Wilson $\chi$PT: because the partially 
quenched quark mass-scale, set by microscopic eigenvalues, is $1/V$, the 
${\cal O}(a^2)$ terms have a large effect even if $a\sim 1/\sqrt{V}$ 
(this is known as the Aoki-regime). Other 
methods to extract the LECs of Wilson $\chi$PT may not be as sensitive 
to the ${\cal O}(a^2)$ if $a\sim 1/\sqrt{V}$. For example, in the 
$p$-regime of Wilson $\chi$PT, where the quark mass scale is $1/L^2$, the 
${\cal O}(a^2)$ may have a much smaller effect if $a\sim 1/\sqrt{V}$ 
(this is known as the GSM regime). For a discussion 
of the counting in the $\epsilon$- and $p$-regime of Wilson $\chi$PT 
and the relation to the Aoki and GSM regimes, see \cite{B2,Shindler:2009ri}.

\section{The theoretical prelude}\label{theory}
\subsection{Twisted mass QCD in the continuum}\label{cont}
The fermionic part of the Lagrangian density of continuum twisted mass QCD, for two flavors, is given, in the twisted basis, by 
\begin{equation}
\mathcal{L}=\bar{\chi}(\gamma_\mu D_\mu+m+iz_t\gamma_5\tau_3)\chi,
\label{contAction}
\end{equation}
where in addition to the usual Dirac term and the quark mass $m$, the so called twisted mass $z_t$ has been introduced. Note that the twisted mass term has a non trivial Dirac and flavor structure, as it comes with $\tau_3$ in flavor space and with $\gamma_5$ in Dirac space. The immediate consequence of the addition of this new mass term is that the determinant of the twisted mass Dirac operator $D$ is strictly positive and one does not encounter the so-called exceptional configurations (these are configurations where the Dirac eigenvalue is almost equal to minus the quark mass and which correspond to an almost singular Dirac operator). This is achieved, since the spectrum of the Dirac operator is excluded by a strip of width $2z_t$ along the real axis~\cite{GS}, but can also immediately be seen from the fact that 
$\det(D+m+iz_t\gamma_5\tau_3)=\det(D+m)\det((D+m)^{\dagger})+z_t^2>0$.
The connection of twisted mass to ordinary QCD is straightforward in the continuum if one 
considers the following chiral transformation~\cite{Frezzossi},
 \begin{equation}
\psi=\exp(i\omega\gamma_5\tau_3/2)\chi, \quad \quad
\bar{\psi}=\bar{\chi}\exp(i\omega\gamma_5\tau_3/2),
\end{equation}
where $\omega=\arctan(z_t/m)$.
Then, one can immediately rewrite the twisted mass Lagrangian density as 
\be
\mathcal{L}=\bar{\psi}(\gamma_\mu D_\mu+M)\psi,
\label{trafocontAction}
\ee
where $M=\sqrt{m^2+z_t^2}$ is the polar mass. The Grassmann fields $\psi$ are in the physical basis. Since the transformation between the two bases is non-anomalous~\cite{Frezzossi}, one can consider it as merely a change of variables which relates twisted mass QCD to ordinary QCD. On the lattice, it was shown in Ref.\ \cite{FrezzottiA} that this equivalence is still valid, but spoiled, as anticipated, by discretization errors. 

\subsection{Wilson $\chi$PT for twisted mass fermions}\label{TMWchPT}

Entirely relying on symmetry properties, one can write down the chiral Lagrangian with $\mathcal{O}(a^2)$ terms included~\cite{SW, SS, RS, BRS}. In this study, we focus on the $\epsilon$-regime, where $m\sim z_t\sim a^2\sim 1/V$, and hence the pion Compton wavelength is much larger than the box where the theory is regulated. Consequently, the partition function factorizes and is given by a zero dimensional unitary matrix integral describing the zero momentum modes \cite{DSV, ADSV, KVZprl, KVZprd, TMSV, SVdyn}. 
The partition function at a fixed vacuum angle $\theta$ is decomposed according to 
\be Z_{N_{\rm f}}(m,\theta;a)=\sum_{\nu=-\infty}^{\infty} e^{i\nu\theta}Z^\nu_{N_{\rm f}}(m;a)
\ee
into an infinite sum of partition functions with a fixed index $\nu$. 
The fixed index partition function for twisted mass fermions with all leading order (LO) in $a$ discretization errors, in the $\epsilon$-regime, reads 

\begin{align}
\label{fullZ}
Z^\nu_{N_{\rm f}}(m)=\int_{\U(N_{\rm f})}
 &d\mu(U)\, {\det}^{\nu}U\exp\left[\frac{m}{2}V\Sigma  \Tr\; (U+U^{-1})+\frac{z}{2}V\Sigma  \Tr \tau_3(
 U-U^\dagger)
\right]&\nonumber\\
&\times\exp\left[ -a^2 VW_6\Tr^2( U+U^{-1})-a ^2VW_7\Tr^2 (U-U^{-1})\right.&\nonumber\\
&\left.-a ^2 VW_8\Tr (U^2+U^{-2})\right],&
\end{align}
where the complex matrix valued spurion fields $m$, $a$ are taken to be real and proportional to the identity. As stated previously, $\nu$ is the index of Wilson Dirac operator 
(defined via the spectral flow lines \cite{SmitVink, Itoh, EHN}).
Note that the partition function, apart from the chiral condensate $\Sigma$, involves three new unknown LECs $W_{6/7/8}$, which parametrize the discretization errors.
Note that in this article, we follow the sign conventions of~\cite{TMSV} which are the opposite of~\cite{BRS}, where the same LECs are given by $-W'_{6/7/8}$, respectively. 
The values for the LECs $W_{6/7/8}$ are determined by the lattice action (e.g.\ a particular choice of the gauge action,
improvement terms and/or the gauge field smearing in the Dirac operator, etc.) and can be determined through lattice simulations. 

 \subsection{The microscopic spectral density for $N_{\rm f}=2$ twisted mass fermions}\label{TManalytical}

 In Ref.~\cite{TMSV}, the microscopic spectral density of the Hermitian Dirac operator $D_5(m=0)\equiv\gamma_5 D(m=0)$ for a fixed index $\nu$ and two flavors at maximal twist was derived analytically in the framework of Wilson $\chi$PT for twisted mass fermions (Wtm$\chi$PT). The authors of~\cite{TMSV} employed the graded method, where one adds an additional fermionic quark and an additional bosonic (ghost) quark with twisted masses $z$ and $z'$, respectively, to the partition function. This prescription is often referred to as partial quenching. In the approximation where $W_6=W_7=0$, the supersymmetric partition function takes the form 
\begin{equation}
Z^\nu_{3|1}({\cal Z};a) = \int_{Gl(3|1)/U(1)} \hspace{-0.5mm}  d\mu(U) \
 {\rm Sdet}(iU)^\nu\;
  e^{\frac{i}{2}{\Str}({\cal Z}[U+U^{-1}])
    + \hat{a}^2{{\Str}(U^2+U^{-2})}},
     \label{gradedZ}
 \end{equation}
 where ${\cal Z}$ contains the appropriate sources with respect to which one differentiates in order to compute the desired spectral resolvent, see~\cite{TMSV}.  Here and below, we will use the 
 notation $\hat{m}=mV \Sigma$, $\hat{z}_t=z_tV\Sigma$ and $\hat{a}^2=a^2VW_8$. 
 Neglecting $W_{6/7}$ is an approximation performed in order to simplify the analytical computation of the integral over the graded group $Gl(3|1)/U(1)$. It is also motivated by the conventional lore that the double-trace terms are suppressed in the large $N_{\rm c}$ limit \cite{KaiserLeutwyler}.
  
 Here, we will state the results from~\cite{TMSV} needed in the present context. The spectral density can be computed through the discontinuity of the resolvent,
\begin{equation}
\rho^\nu_5(\hl^5,\hz_t;\ha) = \left \langle \sum_k \delta(\hl^5_k-\hl^5) \right \rangle_{N_{\rm f}=2} = \frac{1}{\pi}{\rm Im}[G^\nu_{3|1}(\hz=-\hl^5,\hz_t;\ha)]_{\epsilon\to0 },
 \label{rho5def}
 \end{equation}
where $\hl^5=\lambda^5 V\Sigma$ are the rescaled eigenvalues of the Hermitian Wilson Dirac operator. 
One should note that the quark mass scale of the partially quenched flavors is set by the magnitude of the Dirac
eigenvalue we consider. As we will be focusing on the microscopic eigenvalues, the partially quenched flavors automatically have quark masses of order $1/V$.

\begin{figure}[p!]
\begin{center}
\vspace*{-0.55cm}
\includegraphics[height=10cm, angle =-90]{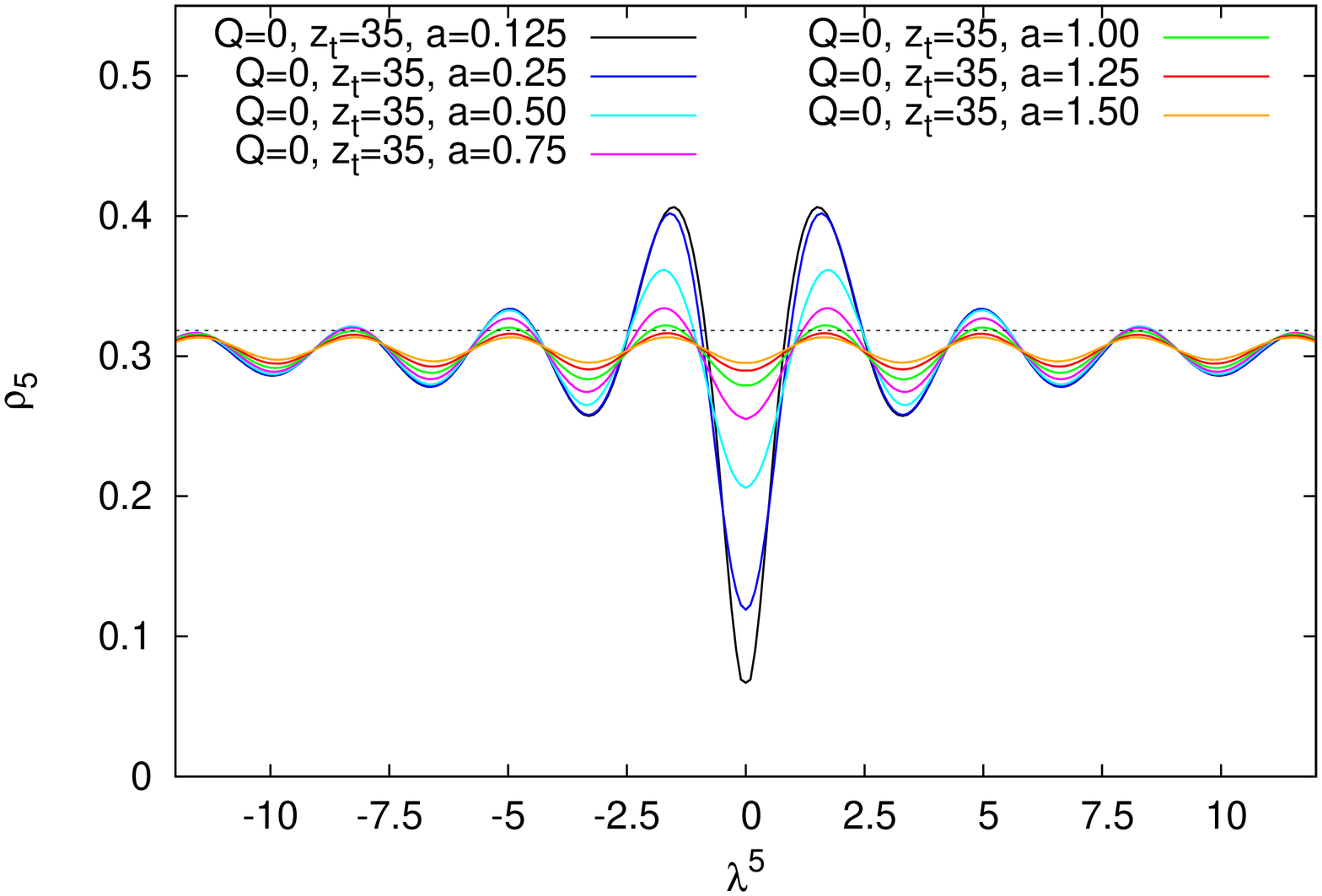}
\vspace*{-0.3 cm}
\includegraphics[height=10cm, angle =-90]{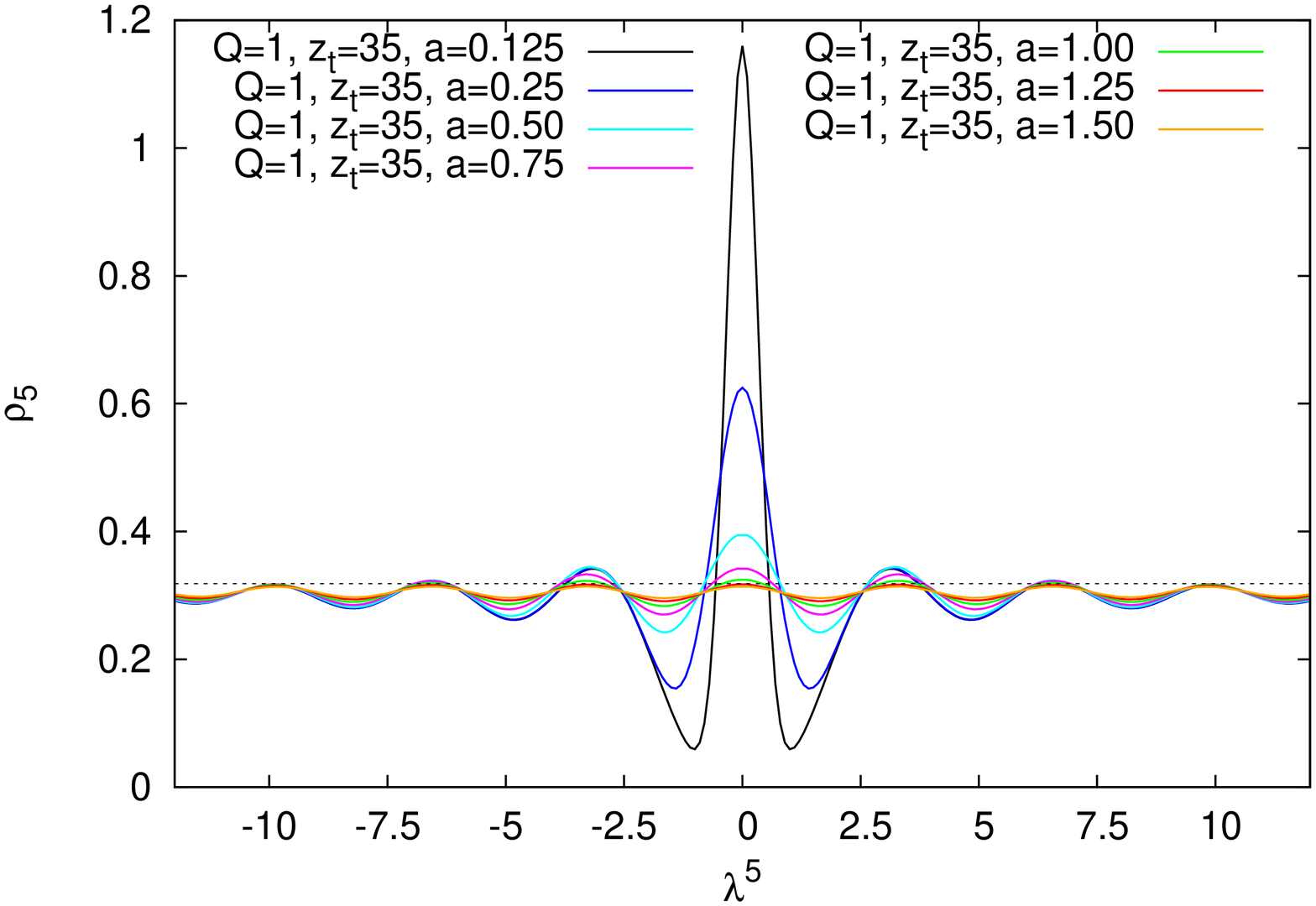}
\vspace*{-0.3 cm}
\includegraphics[height=10cm, angle =-90]{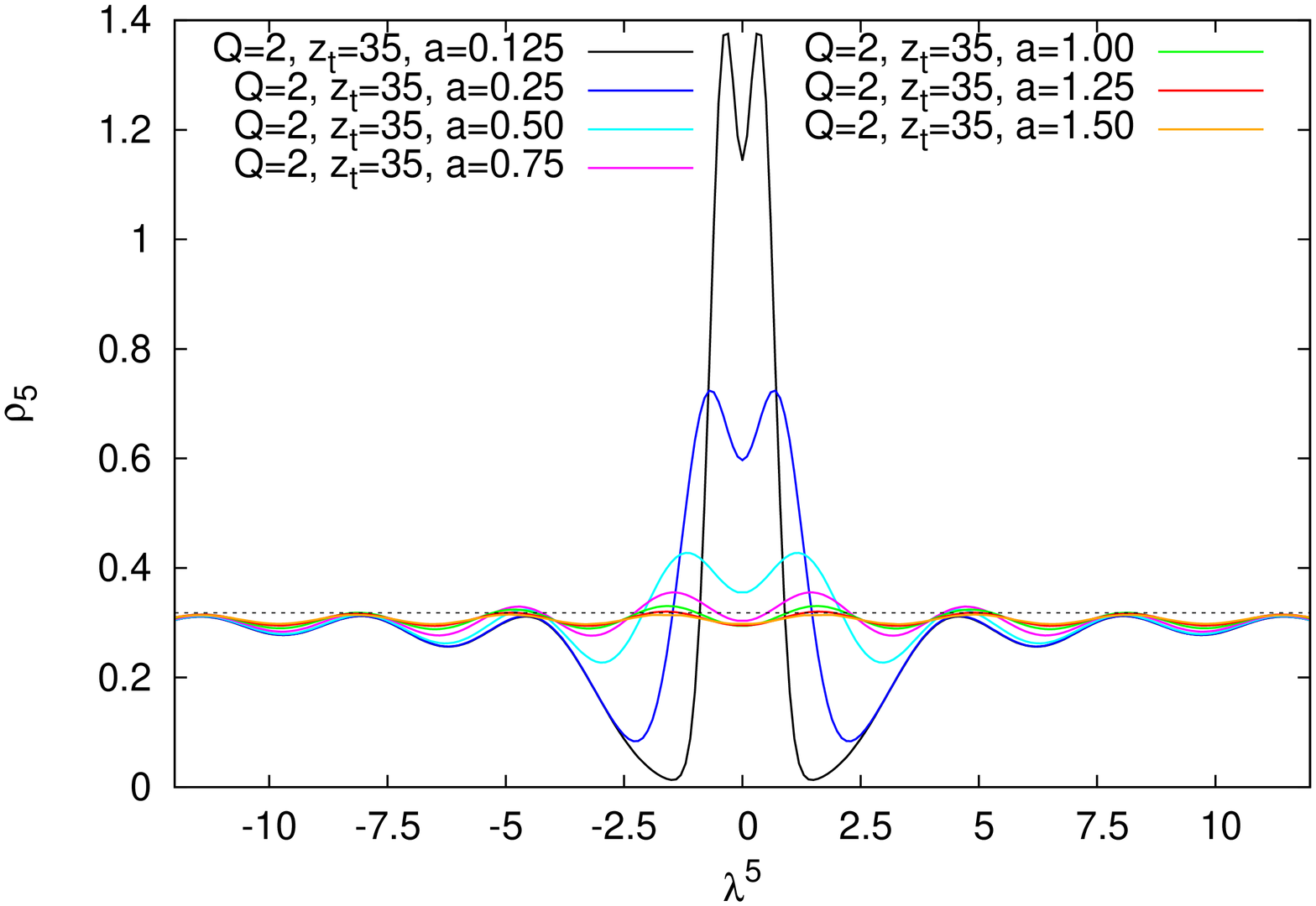}
\end{center}
\vspace*{-0.75cm}
\caption{\label{figz35} \linespread{1.23}\selectfont{The $\epsilon$-regime spectral density of the Hermitian twisted mass Wilson Dirac operator is plotted for the lowest values of the topological charge ($|\nu|=0, 1, 2$ in the first, second and third row, respectively). This is the analytical result derived in \cite{TMSV} plotted for $\hat{z}_t=35$ and different values of the rescaled lattice spacing $\hat{a}$. We see that the microscopic eigenvalue distribution and in particular the would be zero modes, corresponding to the single peak for $|\nu|=1$ and double peak for $|\nu|=2$, are extremely sensitive to the $\mathcal{O}(a^2)$ effects even though $a\sim1/\sqrt{V}$. Note that we have dropped the ``hat'' from the rescaled variables $a$, $z_t$ in the legend of the plots. }
}
\end{figure}

\begin{figure}[p!]
\begin{center}
\vspace*{-0.35cm}
\includegraphics[height=10cm, angle =-90]{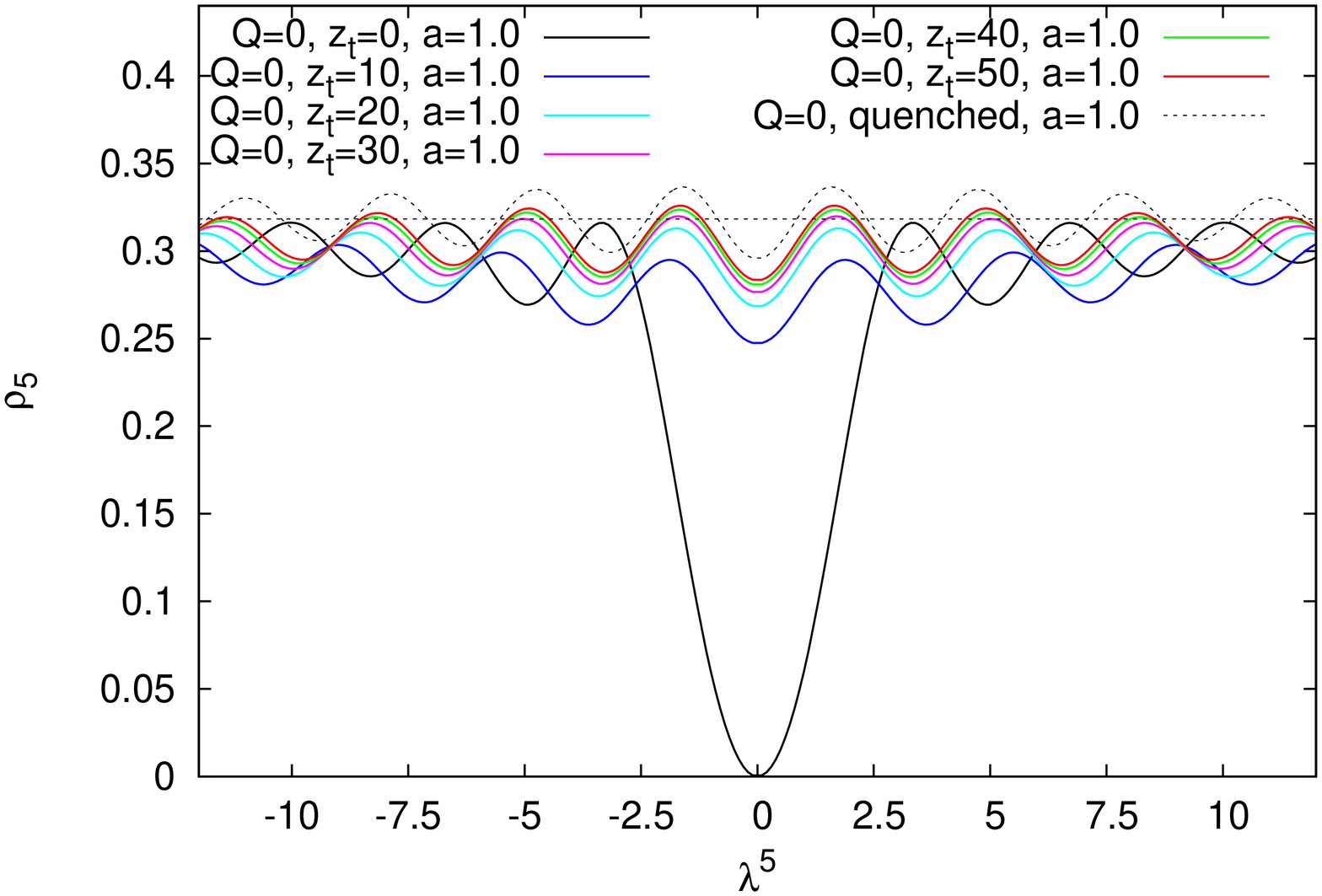}
\vspace*{-0.3 cm}
\includegraphics[height=10cm,  angle =-90]{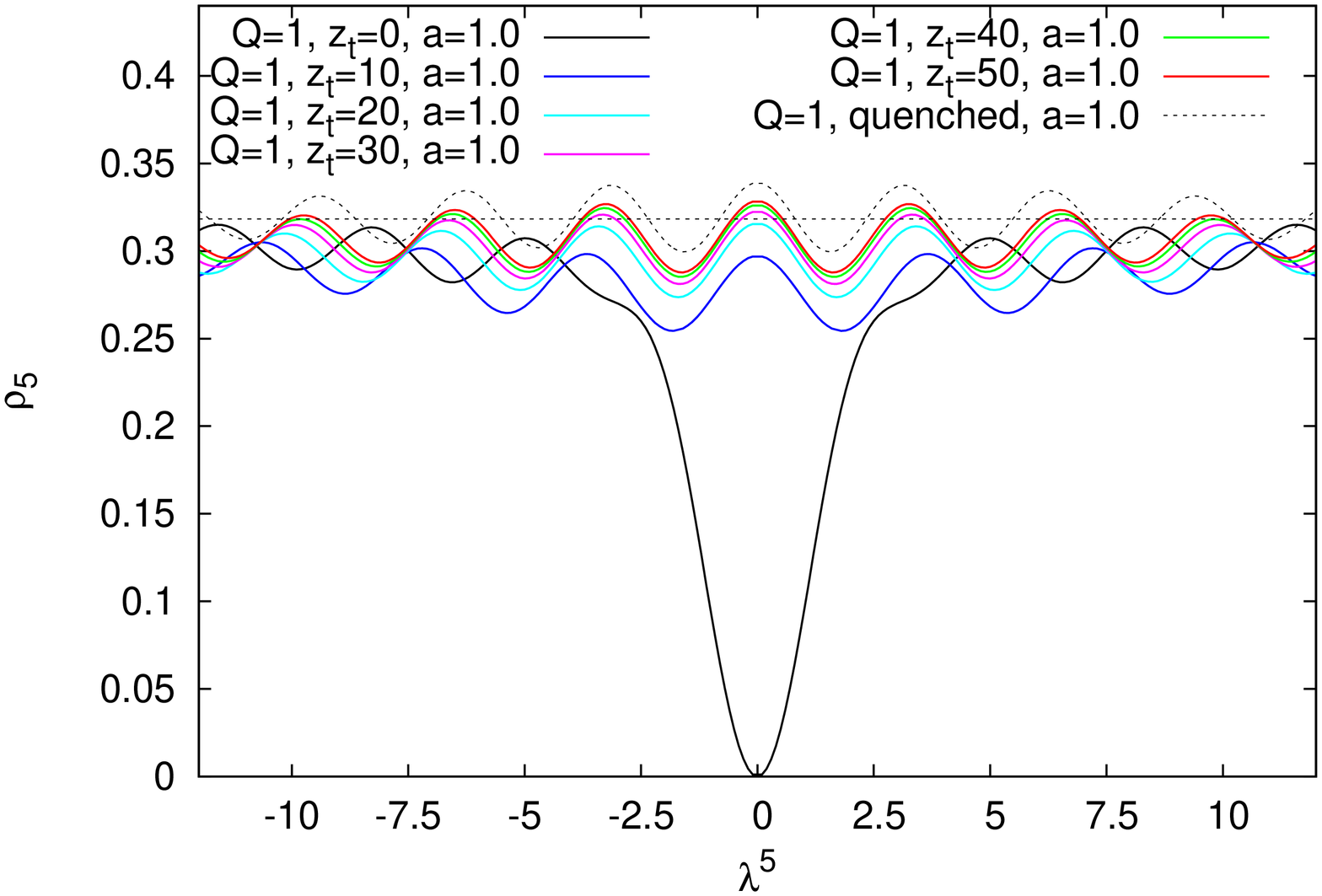}
\vspace*{-0.3 cm}
\includegraphics[height=10cm, angle =-90]{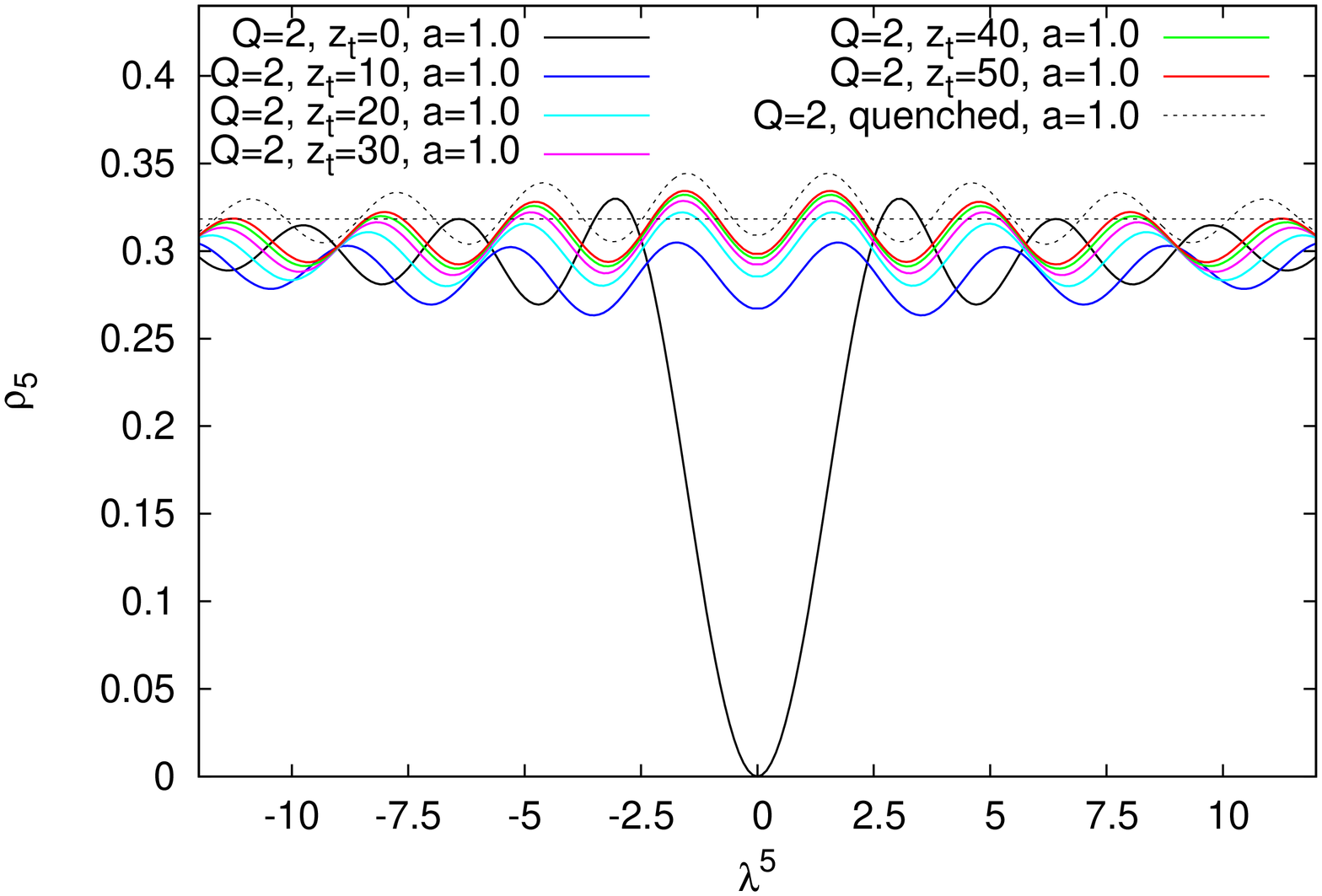}
\end{center}
\vspace*{-0.65cm}
\caption{\label{figa1} As in Figure \ref{figz35}, we plot here the $\epsilon$-regime spectral density of the Hermitian twisted mass Wilson Dirac operator for $|\nu|=0, 1, 2$ obtained in \cite{TMSV}. Here we have chosen $\hat{a}=1$ and we vary $\hat{z}_t$. We include also the quenched spectral density for comparison.
}
\end{figure}

 After a lengthy and technical computation presented in~\cite{TMSV}, the final expression for the resolvent is 
 \begin{align}
  \label{G31factorized}
  G_{3|1}^\nu(z,z_t;a) =
  G_{1|1}^\nu(z,z;a) &+ \frac{Z_2(iz_t,z;a)}{Z_2^\nu(iz_t,-iz_t;a)}\;
  \frac{z-iz_t}{2iz_t}\; G_{1|1}^\nu(-iz_t,z;a)& \nonumber\\
  & - \frac{Z_2^\nu(-iz_t,z;a)}{Z_2^\nu(iz_t,-iz_t;a)}\;
  \frac{z+iz_t}{2iz_t}\; G_{1|1}^\nu(iz_t,z;a),&
 \end{align}
  where
 \begin{align}
  G_{1|1}^\nu(z_1,z_2;a) =  & -\frac{1}{16 a^2 \pi} \int_{-\infty}^\infty dsdt \
     \frac{1}{t+z_2-is-z_1} e^{-(s^2 + t^2)/(16 a^2)}&
  \nonumber \\
  &  \times \left(\frac{is+z_1}{t + z_2}\right)^\nu
  Z_{1|1}^\nu(\sqrt{-(is + z_1)^2},\sqrt{-(t + z_2)^2},a=0),&
  \label{G11gen}
 \end{align}
  with
  \begin{equation}
  Z_{1|1}^\nu(m_1,m_2;a=0) = \left(\frac{m_2}{m_1}\right)^\nu (I_\nu(m_1) m_2 K_{\nu+1}(m_2)+m_1 I_{\nu+1}(m_1)K_{\nu}(m_2))
  \end{equation}
and $I_\nu$ ($K_\nu$) are modified Bessel functions of the first (second) kind.  
The remaining integrals are evaluated numerically in order to produce plots of the spectral density. 
In Figure~\ref{figz35} and~\ref{figa1}, we plot the spectral density of the Hermitian twisted mass Wilson Dirac operator for the range of the parameters which are relevant to this study. 

Figure~\ref{figz35} shows the analytical result of the microscopic ($\epsilon$-regime) spectral density for a fixed large value of the twisted mass ($\hat{z}_t=35$) for various values of the rescaled lattice spacing $\hat{a}$ and for the three values of the index considered in this study ($|\nu|=0, 1, 2$).  
The characteristic feature of continuum or close-to-continuum $\nu=0$ results is the presence of a deep minimum at $\lambda^5=0$ in the spectral density, with $\rho^5(0)=0$ in the continuum, or a pronounced maximum at $\nu\neq0$ (double maximum for $|\nu|=2$), corresponding to the zero modes or would-be zero modes of the Hermitian Dirac operator, becoming a Dirac delta in the continuum.
Thus, the lattice data can already be visually assessed for the presence of these features to see how important are lattice effects.
Note, however, that the relevant parameter is $\hat{a}^2=a^2VW_8$ -- thus even rather small lattice spacings can be devoid of the discussed features, i.e.\ have the minimum/maximum smeared out, if $W_8$ is large.

In Figure~\ref{figa1}, we plot the analytical result of the microscopic spectral density for a large fixed value of the rescaled lattice spacing ($\hat{a}=1$) for several values of the rescaled twisted mass (and for the quenched case).
Here, a significantly different behavior of the spectral density corresponds to small rescaled twisted masses, with a deep minimum present for $\hat{z}_t\approx1$ and exactly vanishing spectral density at the origin in the massless limit.
However, since $\hat{z}_t=z_tV\Sigma$ and the expected value of the chiral condensate is of order 300 MeV, the relevant regime of rescaled twisted masses is of order $\hat{z}_t\approx35$ at this volume, as the value chosen for showing the $\hat{a}$-dependence in Figure~\ref{figz35}.
We note also that the distance between peaks (minima or maxima) of the spectral density is affected by the presence of dynamical light quarks.
In the quenched case, this distance is $\pi/2$, while for finite $\hat{z}_t$ it slightly increases for the low-lying eigenvalues.
Since the area below the spectral density curve is normalized to the number of eigenvalues, a consequence of increased distance between peaks in the dynamical case is that the spectral density lies systematically below $1/\pi$, while the quenched one oscillates about this value.

\section{The computational setup}\label{setup}
\subsection{The lattice action and parameters}\label{Slattice}
In our study, we have been employing gauge field configurations with $N_{\rm f}=2+1+1$ dynamical twisted mass fermions at maximal twist. By courtesy of the European Twisted Mass Collaboration (ETMC), these are publicly available configurations~\cite{Baron1, Baron2}.  In the gauge sector, ETMC employs the Iwasaki 
action~\cite{iwasaki1, iwasaki2}, which is renormalization group improved and reads
\begin{equation}
   \label{eq:Sg}
   S_{\rm gauge} =  \frac{\beta}{3}\sum_x\Biggl(3.648\sum_{\substack{
     \mu,\nu=1\\1\leq\mu<\nu}}^4\{1-\operatorname{Re}\Tr(U^{1\times1}_{x,\mu,\nu})\}\Bigr.
     \Bigl.-0.331\sum_{\substack{\mu,\nu=1\\\mu\neq\nu}}^4\{1
    -\operatorname{Re}\Tr(U^{1\times2}_{x,\mu,\nu})\}\Biggr)\, ,
 \end{equation}
 with $\beta$ the inverse bare gauge coupling, $U^{1\times1}_{x,\mu,\nu}$ is the usual plaquette and $U^{1\times2}_{x,\mu,\nu}$ is the rectangular $(1\times2)$ Wilson loop. In the fermionic sector, we have two variants of the twisted mass action, one for the light degenerate $u$, $d$ quarks and one for the heavy non-degenerate doublet of $s$, $c$ quarks.
The fermionic action for the degenerate light flavors reads \cite{FrezzottiA,Frezzossi,Frezzossi2}
  \begin{equation}
  S_{\rm light}[\chi, \bar{\chi}, U] = \sum_x \bar\chi_l(x) \left[D_W {+} m_{(0,l)} {+} i\gamma_5\tau_3\mu_l \right]\chi_l(x)\,,
  \end{equation}
where $m_{(0,l)}$ is the untwisted bare light quark mass, $\mu_l$ is the bare twisted mass in the light sector.
For the heavy non-degenerate strange and charm quarks, we have \cite{Frezzossi2,Frezzotti:2003xj}
\begin{equation}
    \label{eq:sf}
   S_{\rm heavy}[\chi, \bar{\chi}, U] =\ a^4\sum_x\left\{ \bar\chi_h(x)\left[ D[U] + m_{(0,h)} +
     i\mu_\sigma\gamma_5\tau_1 + \mu_\delta \tau_3 \right]\chi_h(x)\right\}\,,
\end{equation}
where $m_{(0,h)}$ is the untwisted bare quark mass for the heavy
 doublet, $\mu_\sigma$ the bare twisted mass of the heavy doublet.
 Note that the twist angle is this time along the $\tau_{1}$ direction and
 $\mu_\delta$ the mass splitting along the $\tau_{3}$ direction.
The Wilson Dirac operator $ D_{\rm W}$ is defined through the addition of the lattice Laplacian to the naive symmetric covariant derivative
\begin{equation}
 D_{\rm W}=\frac{1}{2}\gamma_\mu(\nabla_{\mu}+\nabla_{\mu}^\star)-\frac{a}{2}\nabla^\star_\mu\nabla_\mu\,,
\end{equation}
where $\nabla_\mu$ and $\nabla_\mu^\star$ denote the forward and backward covariant derivatives.

Our study employs one lattice ensemble with lattice spacing $a=0.0815(30)$ fm \cite{italiani}, the lattice volume is $32^3\times 64$, the physical extent of the box is $L\sim 2.5$ fm and the bare twisted masses are equal to $a\mu_l=0.0055$,  $a\mu_{\sigma}=0.135$ and 
$a\mu_{\delta}=0.170$. 
This ensemble has $2+1+1$ flavors, but the heavy charm and strange quarks, whose bare twisted masses ($a\mu_s=0.0158$ and $a\mu_c=0.2542$) are much larger than the smallest Dirac eigenvalues, behave as quenched from the point of view of the Dirac spectrum. This allows us to compare with the analytical results of the spectral density with $N_{\rm f}=2$.
The pion mass ($m_{\pi}$) computed from these configurations is equal to around $370$ MeV.  The quite large physical extent of the box takes care of the finite volume corrections, which are $\mathcal{O}(e^{-m_{\pi}L})$, and $m_{\pi} L\approx 5$ for our setup.  

Even though the pion mass in this simulation does not satisfy $1/M_{\pi}\gg L$, the smallest eigenvalues can be in the $\epsilon$-regime. The characteristic energy scale below which the Dirac eigenvalues are described by the $\epsilon$-regime of $\chi$PT is called, in an analogy to the condensed matter literature, the Thouless energy scale \cite{jamesjac} (see also below). It is important to stress that the number of eigenvalues in the $\epsilon$-regime in a given lattice simulation is not known beforehand.

\subsection{The computation of the index}\label{Qtop}

The integer $\nu$ in Wtm$\chi$PT is the index of the Wilson Dirac operator. 
The direct numerical computation of the index is quite demanding. In our study, we have therefore utilized a gluonic definition of the topological charge combined with smearing via the Wilson flow \cite{WF}. The Wilson flow is an economical method (with respect to the direct computation of the index utilizing the overlap Dirac operator), with solid theoretical foundations. Another attractive feature of it is that it does not involve additive or multiplicative renormalization. However, it is important to point out that any gluonic definition that amounts to computing the volume integral of the topological charge density of smoothed configurations does not yield an integer number in general.

In our study, we have associated an integer topological charge $\nu$ to gauge configurations which had a measured topological charge in the range $[\nu-1/2, \nu+1/2)$. Defining the topological charge in this manner is by no means a caveat of our analysis, since at finite lattice spacing the topological charge is not well defined and there is no possibility of a unique assignment of a particular value of the topological charge to a lattice gauge configuration. Even if one uses 
a costly fermionic method that yields an integer value of the topological charge, such as the index of the overlap Dirac operator, still the definition is not unique, because the value of the index is dependent on the $s$ parameter of the kernel of the overlap operator that has to be tuned appropriately in order to ensure locality \cite{local,Krzmixed}.
In order to test the sensitivity to the used method, we have employed various discretizations of the topological charge density. The first discretization as mentioned, utilizes the Wilson plaquette definition and has discretization errors of $\mathcal{O}(a^2)$~\cite{marga}, the second definition includes the addition of the clover term and has discretization errors of $\mathcal{O}(a^2)$~\cite{marga}. The third discretization has rectangular clover terms and has discretization errors of $\mathcal{O}(a^4)$~\cite{marga}. The agreement and correlation among these three methods for the given value of the lattice spacing is above $98\%$, see \cite{andreas,Alexandrou:2017hqw} for a more detailed analysis and discussion of correlations between different definitions of the topological charge. 

We computed the topological charge for 5000 independent gauge field configurations (sufficiently separated in Monte Carlo time such that no autocorrelations are detectable) and selected the ones with index $|\nu|=0,1,2$.
The number of configurations in the three topological sectors turned out to be $\sim 200$ for $\nu=0$, $\sim 400$ for $|\nu|=1$ and also $\sim 400$ for $|\nu|=2$.
Hence, only around 1000 configurations could be used in our analysis, since for this large physical volume, the fluctuations of the topological charge are relatively large.
Note that this is already very many configurations for the standards of present-day large-scale lattice QCD simulations, since this ensemble is the ETMC's longest ensemble.
However, the number of configurations that can be used for our study is naturally limited by autocorrelations in the topological charge and the requirement of analyzing configurations at a fixed index.

\section{Analysis strategy and results}\label{analysis}
\subsection{Analysis strategy}\label{strategy}
We computed the five lowest eigenvalues of the Hermitian operator $D^\dagger D$ at a fixed index $|\nu|=0,1,2$. 
To compare with the analytical formulae for the operator $\gamma_5 D$, we took the square root of the eigenvalues of $D^\dagger D$ and we also determined the sign of each eigenvalue by applying the operator $\gamma_5 D$ to the computed eigenvectors.
In this way, we arrived at five lowest eigenvalues of $\gamma_5 D$, which we denote by $\lambda^5$.
We also checked that the obtained spectra are indeed symmetric with respect to the change $\lambda^5\rightarrow-\lambda^5$ and we symmetrized the spectrum to effectively gain statistics.
Finally, we constructed histograms of eigenvalues for all three topological sectors by selecting a bin size $\delta$ and attributing the eigenvalues with $a|\lambda^5|\in[n\delta,(n+1)\delta]$ to the $n$-th bin ($n=0,\,1,\,2,\,\ldots$).
To estimate the potential systematic effect from choosing a particular bin size, we repeated all our analyses for $\delta=0.8\cdot10^{-4},\,1.0\cdot10^{-4},\,1.2\cdot10^{-4},\,1.4\cdot10^{-4}$.

We performed a bootstrap procedure with 1000 samples to obtain also the errors of the histograms and thus also of the spectral density that can be fitted to Wtm$\chi$PT predictions.
At this stage, the eigenvalues are not yet rescaled $\lambda^5\rightarrow\hl^5=\lambda^5 V\Sigma$, since $\Sigma$ is to be extracted from our fits.

The first thing to address in the analysis of the obtained histograms is whether the data are indeed in the $\epsilon$-regime.
We will discuss this issue in the next subsection by considering the Thouless energy scale and by comparing results of fits assuming either $\epsilon$- or $p$-regimes.
Now, we describe our fitting procedure for the $\epsilon$-regime case.

The fitting ansatz is given by
\begin{equation}
\rho^\nu_5(N\hl^5)^{\rm lattice}=\rho^\nu_5(\hl^5,\hz_t;\ha),
 \label{ansatz} 
\end{equation}
where the right-hand side $\rho^\nu_5(\hl^5,\hz_t;\ha)$ is the analytical formula given by Eq.~(\ref{rho5def}) and the left-hand side are the lattice data ($N$ is a fitting parameter whose role is explained in what follows).
There are, thus, three fitting parameters: the rescaled variables $\hz_t$, $\ha$ and a third parameter that needs to be introduced for a proper description of data.
It is clear that the approximation of neglecting the terms proportional to $W_6$ and $W_7$ in the chiral Lagrangian can only be justified in the large-$N_c$ limit, where single trace terms dominate. This is by far not the case here and this can only be treated as an approximation that simplifies the cumbersome analytical solution. 
In Ref.~\cite{KVZprd}, the exact analytical dependence on $W_6$ and $W_7$ was studied for all the different eigenvalue densities (complex, real) of the unimproved non-Hermitian Wilson Dirac operator. In Figure 1 of~\cite{KVZprd}, the effect of these LECs is described schematically. What was observed was that $W_6$ leads to a broadening of the Dirac spectrum parallel to the real axis according to a Gaussian with a width proportional to $\hat{a}_6=\sqrt{V W_6}a$. Also $W_7$ has a non-trivial effect on the spectrum of the unimproved Wilson Dirac operator and once $W_6 = 0$, the purely imaginary eigenvalues enter the real axis via
the origin, while the real eigenvalues are broadened by a Gaussian with a width proportional to $\hat{a}_7=\sqrt{V W_7}a$. In order to take into account this effect, we allow for a free normalization in the $x-$axis (the $N$ parameter), since this accounts to a certain extent for this broadening or squeezing of the spectrum due to $W_6$ and $W_7$. When everything is properly taken into consideration, the locations of the peaks of the spectral density correspond to single eigenvalue distributions and therefore it is clear that lattice and analytical data have to perfectly agree on the location of the peaks. 

The form of the analytical formulae describing the spectral density implies that a standard fitting procedure, evaluating the analytical formulae at each solver procedure iteration, is by far too demanding.
Therefore, our first step was to tabulate the analytical values by evaluating them for each combination $(|\nu|,\,\hat{a},\,\hat{z}_t,\hat{\lambda}_5)$, with $|\nu|=0,1,2$, $\hat{a}=0.25,\,0.30,\,\ldots,\,1.45,\,1.50$, $\hat{z}_t=5,\,5.1,\,\ldots,\,50.8,\,50.9$ and $\hat{\lambda}_5=0,\,0.1,\,\ldots,\,11.9,\,12.0$, making up a total of 4341480 evaluations. Each evaluation required computing six two-dimensional complex (four-dimensional real) improper integrals, with carefully tuned numerical integration ranges to replace infinities with sufficiently high cutoffs.
The chosen cutoff is $\hat{a}$ and $\hat{z}_t$ dependent, with larger cutoffs for larger $\hat{a}$ and $\hat{z}_t$.
The most difficult cases, for $|\nu|=2$ with large $\hat{a}$ and large $\hat{z}_t$, required computations lasting minutes, thus making the tabulation a somewhat tedious procedure, but a necessary one in order to perform the fits in an efficient way.
The integrations were performed using the \emph{cubature} library \cite{cubature} (using adaptive multivariate integration over hypercubes \cite{Berntsen}).
Having the tabulated values of the analytical spectral density, we could use them in the fits.
For any needed value of parameters, our fitting code performed a short interpolation between available values of $(\hat{a},\,\hat{z}_t,\hat{\lambda}_5)$ for a given $\nu$.
In this way, we obtained values of the fitting parameters $\hat{a},\,\hat{z}_t,\,N$ minimizing the $\chi^2$ function, defined in the standard way as the sum of squared differences between lattice data and the analytical formula. 
We note that the independent variable entering the fitting ansatz, $\hl^5=\lambda^5 V\Sigma$, involves the LEC $\Sigma$ that enters also the fitting parameter $\hat{z}_t$.
Hence, the fitting has to be done self-consistently, i.e.\ changing the value of $\hat{z}_t$ in a solver iteration implies also rescaling the $x$-axis.

It is important to emphasize that the bare condensate values extracted from the matching of the analytical formulae to the lattice data require multiplicative renormalization and this is the only renormalization needed in our procedure. For the case of twisted mass fermions, the relevant renormalization function is $Z_P$ (contrasted to $Z_S$ for ordinary untwisted Wilson fermions). $Z_P$ was computed for these values of the parameters by the ETMC in~\cite{italiani, francais, cyprus} and it was found to be, in the $\msb$ scheme at 2 GeV, $Z_P=0.509(4)$ (we use the value from \cite{italiani}).

For comparison, we also performed fits of the $\epsilon$-regime continuum formula (enforcing $\ha=0$), as well as of the LO $p$-regime continuum formula, which reads
\begin{equation}
\rho^\nu_5(\hl^5)^{\rm lattice}=\frac{1}{\pi}.
 \label{banks} 
\end{equation}
Note that the fitting parameter, $\Sigma$, is here hidden in the left-hand side, i.e.\ finding its $\chi^2$-minimizing value consists in adjusting the rescaling $\hl^5=\lambda^5 V\Sigma$ to solve the minimization problem.
We also remark that there exists a NLO $p$-regime formula \cite{Necco:2011vx}, but it is for Wilson fermions, with no twist, and hence it is not applicable here.


\subsection{$\epsilon$-regime vs.\ $p$-regime}\label{epsvsp}
Before we embark on extracting results using the analytic leading order
$\epsilon$-regime results, we discuss here the magnitude of the number of
Dirac eigenvalues expected to be in the $\epsilon$-regime for the
lattice setup used in this paper.

The scale below which the Dirac eigenvalues behave according to the
leading order $\epsilon$-regime results is known as the Thouless energy scale \cite{jamesjac}.
The analysis of~\cite{jamesjac} leads to the following formula
\begin{equation}
\hat\lambda^5_{\rm Thouless}=1/2(F_{\pi}L)^2.
\label{eq:Thouless}
\end{equation}

With the lattice setup used in the current study, we have $L \sim 2.5$ fm and hence $(F_\pi L)^2 \simeq 3$. Thus for the lattice employed $\hat\lambda^5_{\rm Thouless}\approx1.36$.
This is slightly larger than in previous studies, such as \cite{DHS1, DWW, DHS2}. In their case, due to very high statistics that is easily achievable in quenched simulations, very good agreement was found among the low-lying Dirac spectrum and the Wilson-$\chi$PT predictions.

\begin{figure}[p!]
\begin{center}
\vspace*{-0.95cm}
\includegraphics[width=6cm, angle =-90]{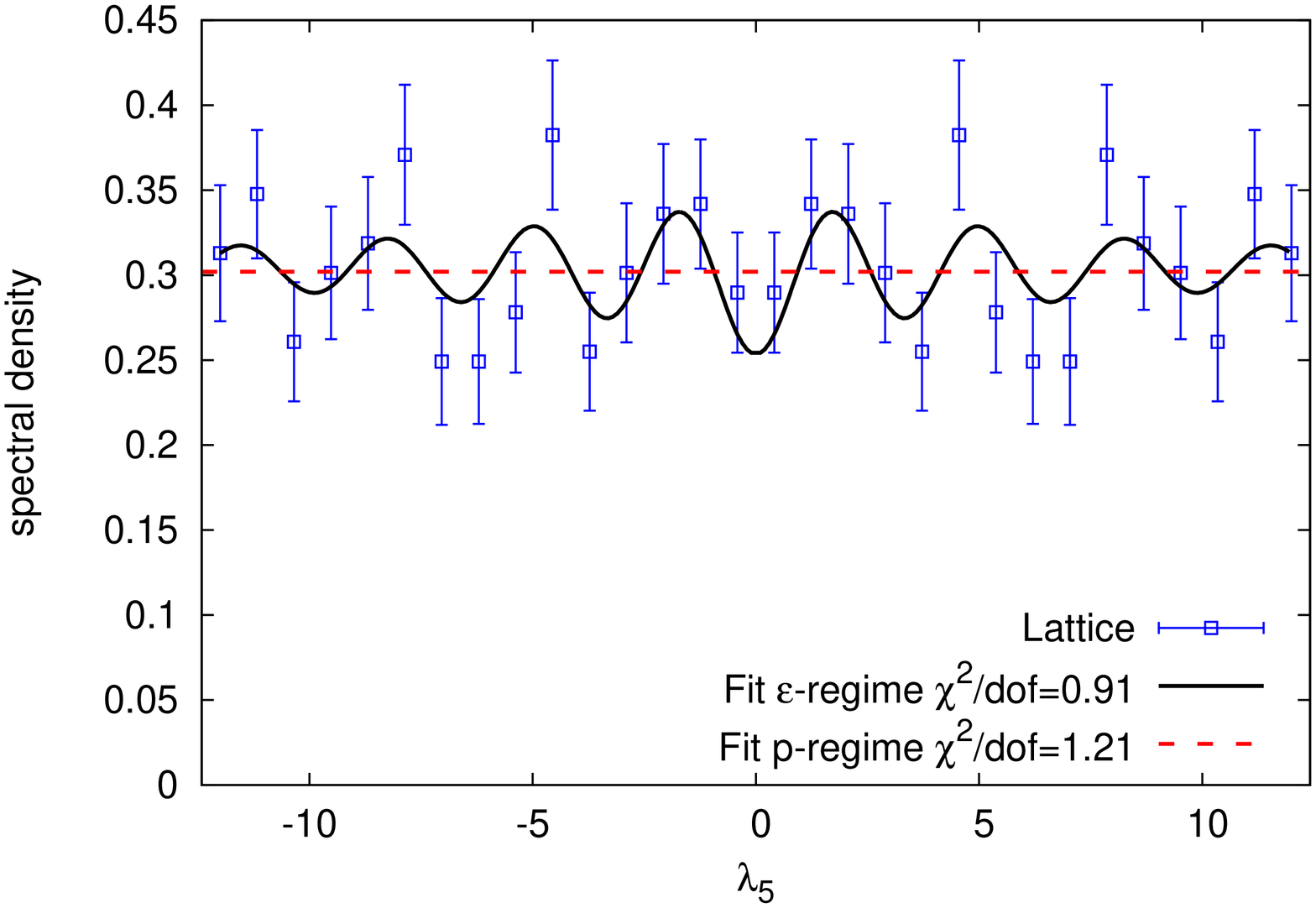}
\vspace*{-0.3 cm}
\includegraphics[width=6cm,  angle =-90]{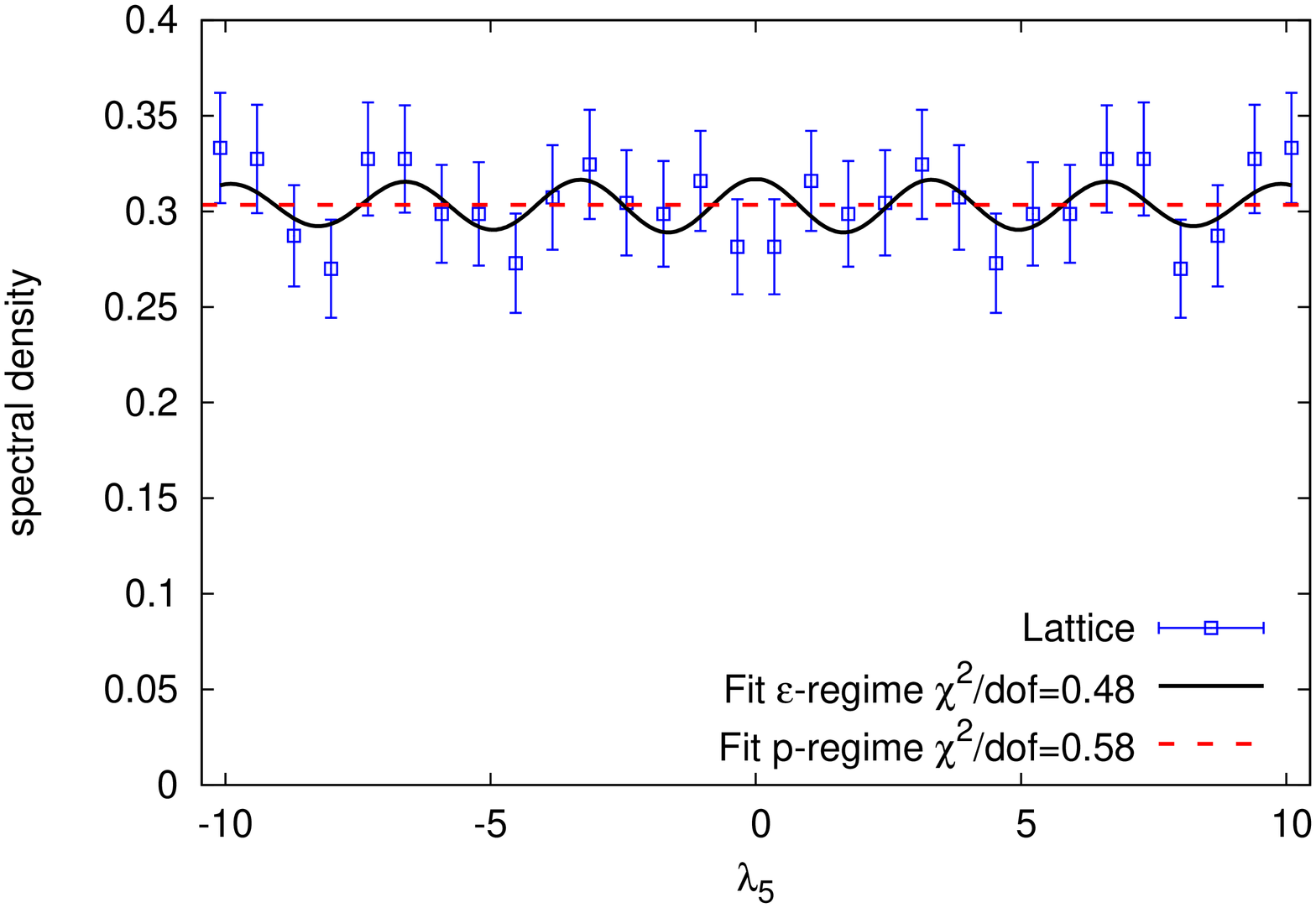}
\vspace*{-0.3 cm}
\includegraphics[width=6cm, angle =-90]{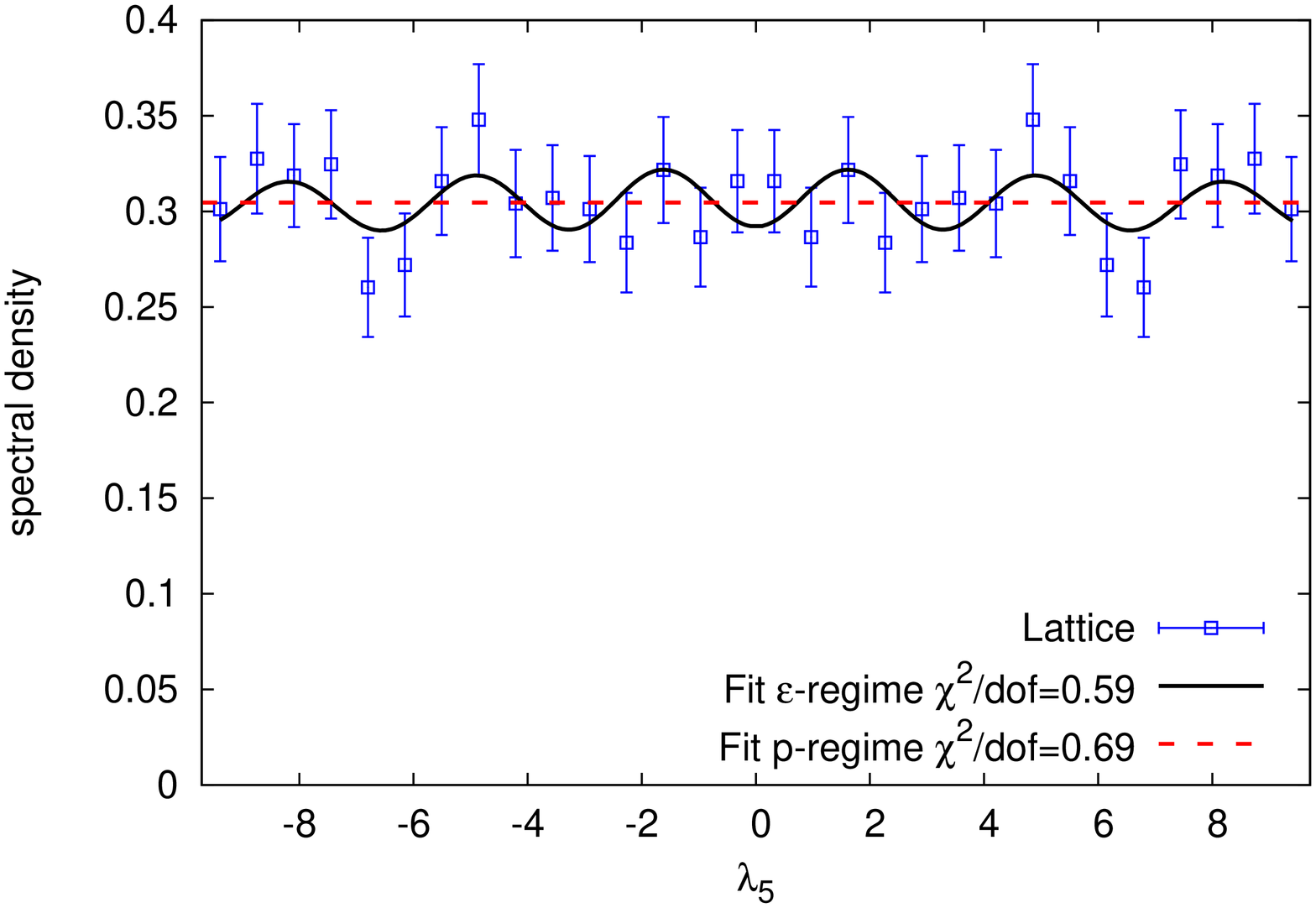}
\end{center}
\vspace*{-0.65cm}
\caption{\label{fig1} The spectral density of the Hermitian twisted mass Wilson Dirac operator is plotted for the lowest values of the index. The solid black curves are fits of the $\epsilon$-regime results, while the dashed red lines are fits of the $p$-regime formula (with additional rescaling for visual purposes, see main text).  The blue data points are the numerical results from a lattice simulation on a $32^3\times 64$ lattice with a lattice spacing $a=0.0815$ fm and twisted mass $a \mu=0.0055$. 
The top plot contains the fitting results of the topological sector with  $\nu=0$, the middle, the results for $|\nu|=1$ and the bottom plot corresponds to $|\nu|=2$. 
}
\end{figure}

In Figure \ref{fig1}, we present the comparison of the fitted analytical results in the $\epsilon$-regime (solid black lines) and the $p$-regime (dashed red lines) vs.\ histograms of lattice data (blue data points with errors from the bootstrap procedure) for the spectral density $\rho^5$ of the Hermitian Wilson Dirac operator $D_5$ in the sectors with $|\nu|=0, 1, 2$. 
Note that the eigenvalue rescaling is different in both regimes, i.e.\ $\hl^5_p:=\lambda^5 V\Sigma_p\neq\lambda^5 V\Sigma_\epsilon=:\hl^5_\epsilon$ (where the subscript $p/\epsilon$ corresponds to fits of the $p/\epsilon$-regime formulae), and hence the plots of $p$-regime and $\epsilon$-regime spectral densitites have different $x$-axes.
To have a comparison in a single plot, we rescaled the $x$-axis of the $p$-regime spectral density by multiplying it by the ratio $\Sigma_\epsilon/\Sigma_p$.
In this way, the $x$-axis of this plot is $\hl^5_\epsilon$ for both spectral densitites.
Obviously, to maintain the correct normalization of the $p$-regime spectral density, we multiplied it by the inverse of the eigenvalue rescaling, i.e.\ by $\Sigma_p/\Sigma_\epsilon$.
Thus, the above mentioned different values of $\Sigma$ extracted in the two regimes are reflected in the positions of the dashed red lines. The ratio of the values in the plots and $1/\pi$ reflects the ratio $\Sigma_p/\Sigma_\epsilon$.
We find that this ratio differs from 1 by 4-6\% for the different topological sectors, which means that the extracted chiral condensate values are consistent with each other when the statistical uncertainty is taken into account.


In general, the fits in the $\epsilon$-regime lead to consistently lower values of the reduced $\chi^2$, especially in the trivial topological sector.
However, the difference is not very large and one can not fully exclude that the eigenvalues are in the $p$-regime.
It is clear that the reason for this are the large statistical errors, even though our statistics is as good as could be achieved with the presently available ensembles.
As we already mentioned, we used the longest ETMC ensemble, but we had to select only configurations with a low topological charge to fit the fixed index formulae \footnote{For $|\nu|\geq3$, the tabulation of analytical values is basically unfeasible, since the integrals to be computed become very difficult and would require computations lasting tens of minutes for each $(\hat{a},\,\hat{z}_t,\hat{\lambda}_5)$.} and moreover we were restricted by autocorrelations.
Thus, the only plausible way to increase statistics would be to generate twisted mass configurations in a fixed topology, with a few thousand independent configurations in each of the three lowest topological sectors.

\begin{figure}[t!]
\begin{center}
\vspace*{-0.35cm}
\includegraphics[height=10cm, angle =-90]{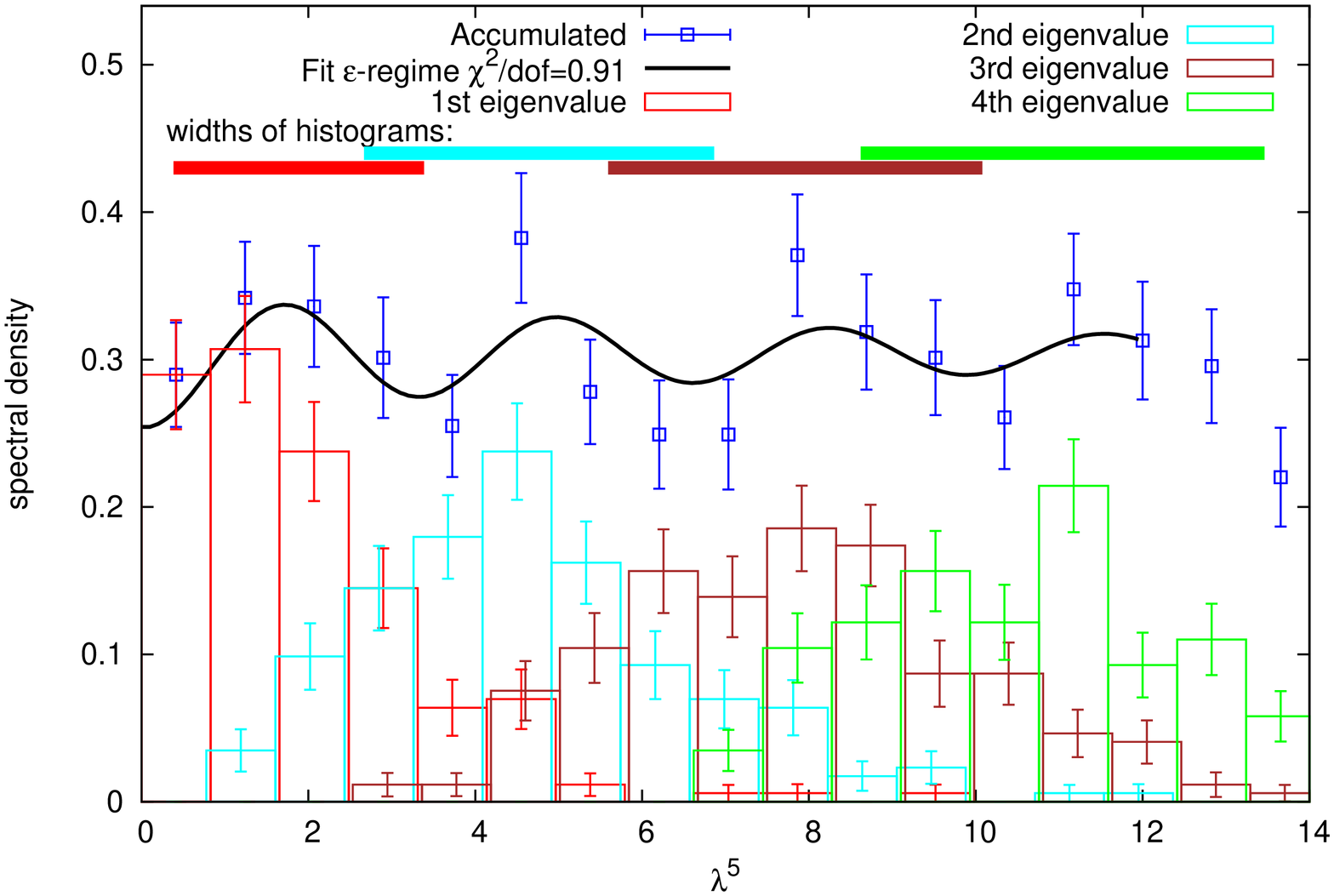}
\end{center}
\vspace*{-0.65cm}
\caption{\label{histo} The spectral density of the Hermitian twisted mass Wilson Dirac operator, together with the $\epsilon$-regime fit and histograms of the 4 lowest eigenvalues. In the upper part of the plot, we show the corresponding widths of the histograms.
}
\end{figure}

We investigate the indications of being in the $\epsilon$-regime further in Figure \ref{histo}.
It shows our $\epsilon$-regime fit ($\nu=0$) together with separate histograms (translated to spectral densities) of the lowest 4 eigenvalues and their widths, defined as standard deviations of their means \footnote{We have also checked an alternative definition of a histogram width, which is the number of eigenvalues contained between the 84th and 16th centiles. For a strictly Gaussian distribution, such definition coincides with the standard deviation. Here, the difference between this definition and the one from the standard deviation is immaterial for our argument and hence we plot only the standard deviations.}.
The characteristic feature of the $\epsilon$-regime is eigenvalue repulsion, i.e.\ only a small overlap between histograms of neighboring eigenvalues, see e.g.\ Figure 3 of Ref.\ \cite{Akemann:2012pn}.
On the other hand, the $p$-regime is characterized by a strong overlap between such histograms, i.e.\ there is no eigenvalue repulsion and the spectral density at a given value of $\hl^5$ comes from tens or hundreds of eigenvalues, as was found numerically in~\cite{KrzSigma, KrzSigma2}.
In our data, we observe only a small overlap between the widths, especially for the lowest two eigenvalues.
In order not to assume what is the number of eigenvalues in the $\epsilon$-regime, our fits were performed for spectral densitites constructed from 1, 2, 3 or 4 eigenvalues, see the next subsection.

All of the above indicates that our $\epsilon$-regime fits are favored with respect to the $p$-regime ones.
It is plausible that already with the 2nd eigenvalue we are entering an intermediate regime between the deep $\epsilon$-regime and the deep $p$-regime.
Nevertheless, the histograms of single eigenvalues are still pretty robustly separated and one can expect that the LO $\epsilon$-regime formula still works considerably good.
Therefore, in the next section we take these fits as our preferred ones and we show the extracted values of LECs.


\begin{figure}[p!]
\begin{center}
\vspace*{-0.95cm}
\includegraphics[width=6cm, angle =-90]{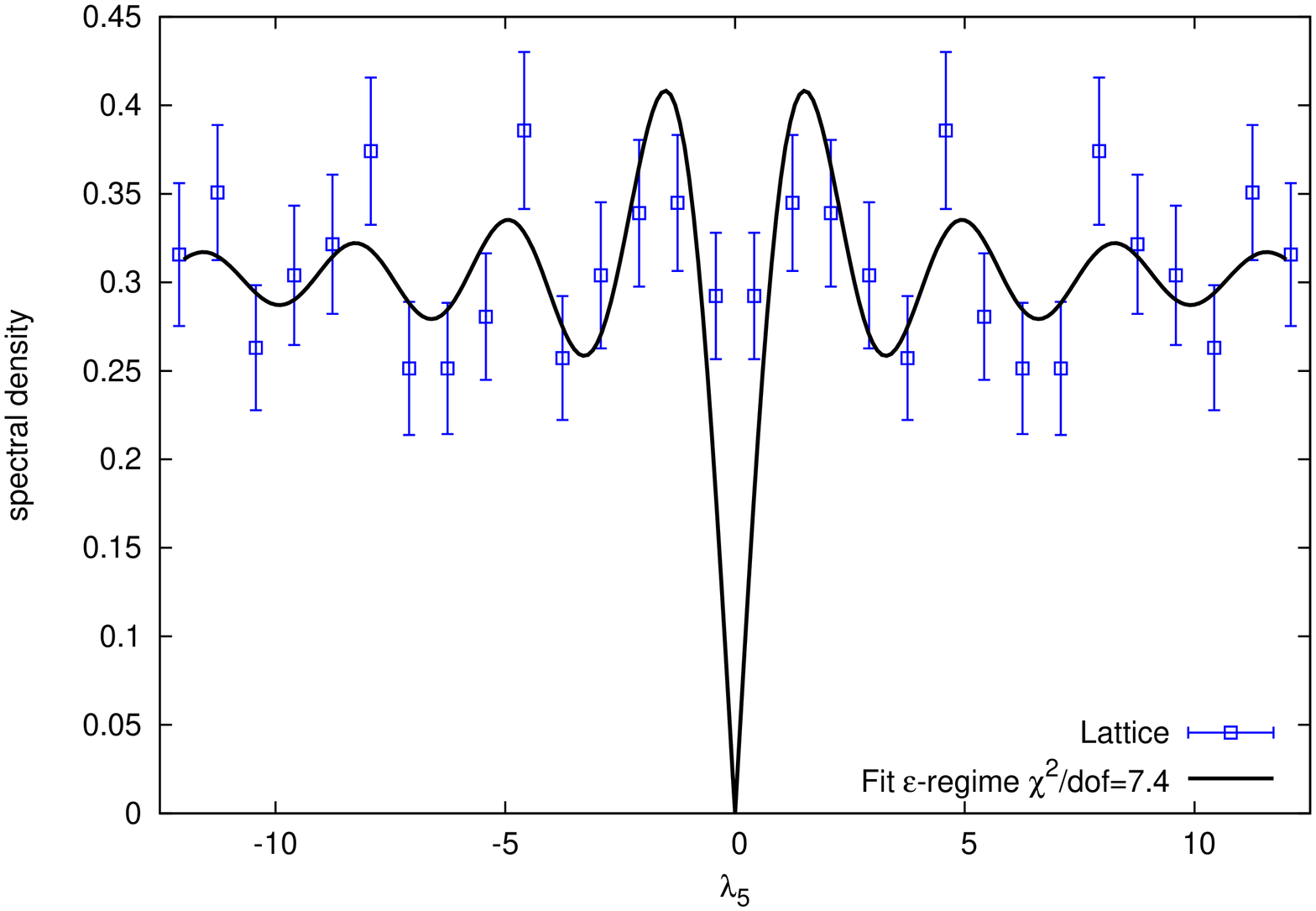}
\vspace*{-0.3 cm}
\includegraphics[width=6cm,  angle =-90]{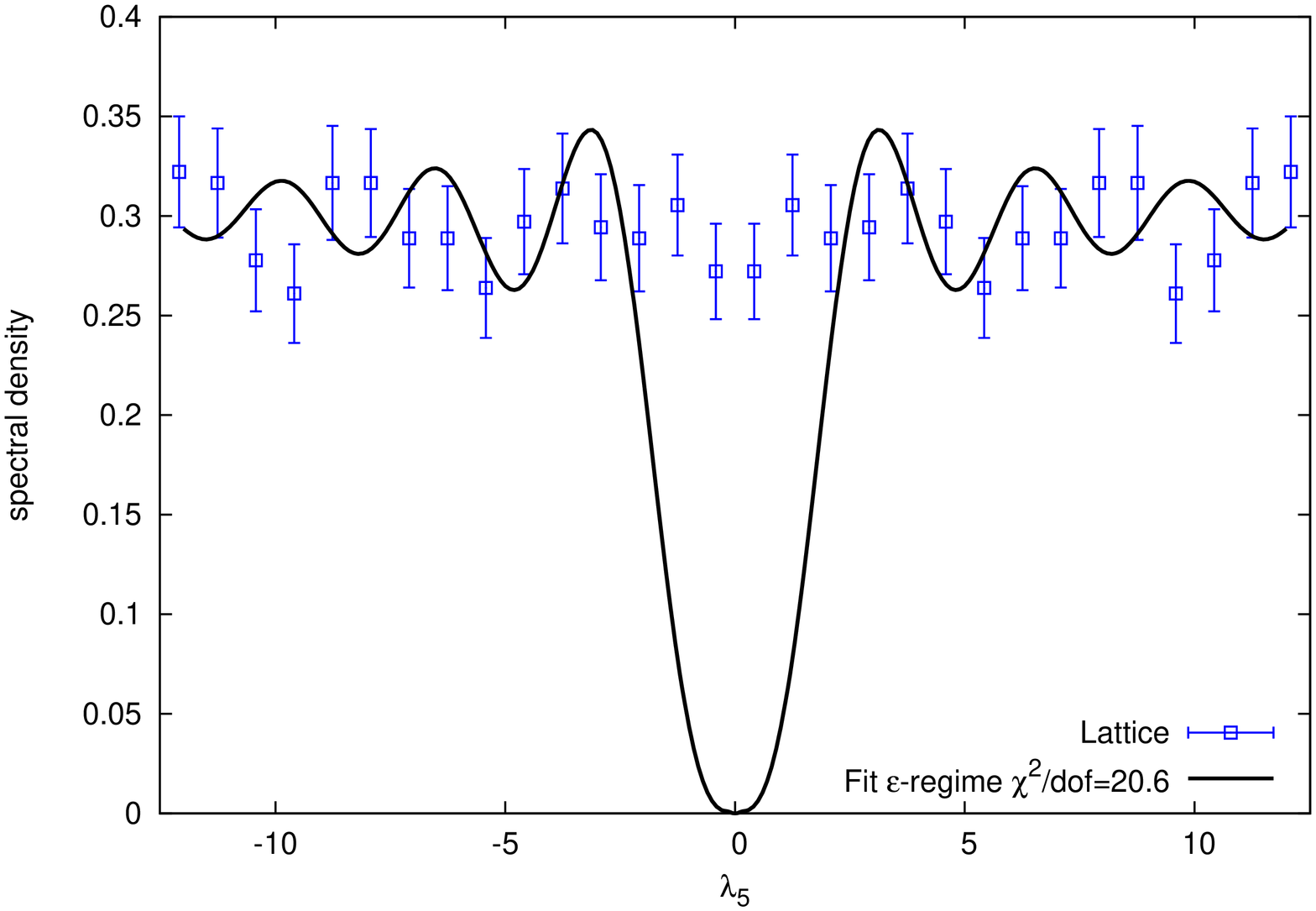}
\vspace*{-0.3 cm}
\includegraphics[width=6cm, angle =-90]{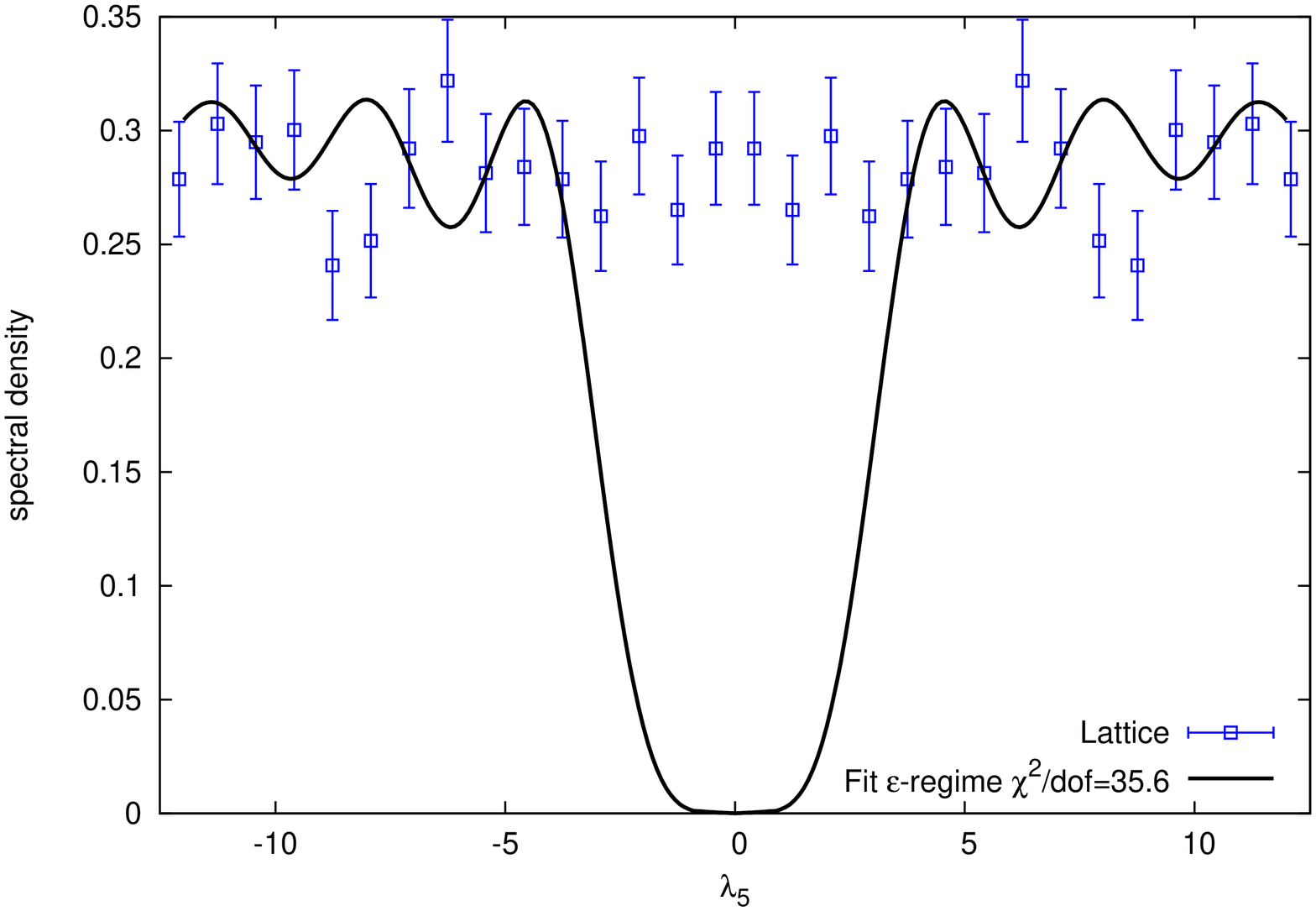}
\end{center}
\vspace*{-0.65cm}
\caption{\label{figcont} The $\epsilon$-regime spectral density of the Hermitian twisted mass Wilson Dirac operator and fits to the continuum analytical results, corresponding to the $a\to 0$ limit of the results derived in \cite{TMSV}. The top plot contains the fitting results of the topological sector with  $\nu=0$, the middle, the results for $|\nu|=1$ and the bottom plot corresponds to $|\nu|=2$. 
}
\end{figure}

\subsection{Extraction of $\Sigma$ and $W_8$}\label{results}
From our fitting parameters $\hz_t$ and $\ha$, we can obtain the LECs $\Sigma$ and $W_8$ through the definition of the scaling variables. 
Before we discuss results from the $\epsilon$-regime fits for twisted mass fermions, we address one more issue that could be potentially raised by the reader, whether the continuum formulae of $\epsilon$-regime $\chi$PT would provide a reasonable description of the data, due to the $\mathcal{O}(a)$ improvement of twisted mass fermions. 
For this purpose we have performed fits with the continuum expression of the microscopic spectral density that can be obtained by the $a\to 0$ limit of Eq.~(\ref{G31factorized}).
These fits are shown in Figure \ref{figcont} and the corresponding values of the fitting parameter $\hz_t$ and the implied values of $\Sigma$ in Table \ref{tabcont}.
The large values of $\chi^2/{\rm dof}$ when neglecting lattice artifacts demonstrate that a continuum fit is not preferred.
It is very important to add that for $\nu\neq 0$ in the continuum due to the Atiyah-Singer index theorem we have exact zero modes (that one can not have on the lattice with any type of Wilson fermions) and thus the corresponding eigenvalue density has Dirac delta peaks at zero.
In our fits, we have explicitly not included the delta functions in the analytic formulae, since the value of $\chi^2/{\rm dof}$ would diverge.
Nevertheless, the behavior of the continuum formulae around $\hl^5=0$ always excludes the continuum formulae, as can be also seen in Figure \ref{figz35} -- in the continuum, or very close to it, there is a deep minimum ($\nu=0$) or a pronounced maximum ($\nu\neq0$) of the spectral density.

\begin{table}[t!]
\begin{center}
\begin{tabular}{ | c || c | c | c | }
   \hline $|\nu |$ & 0 & 1 &2 \\ \hline\hline
   $\hat{z}_t$ &38.25(88)(3.1) & 38.25(39)(2.2)&38.25(92)(3.1)\\ \hline
   $\Sigma^{1/3}$\,\,[MeV] &288.3(2.2)(7.5) & 288.3(1.0)(6) &288.3(2.3)(7.5)\\ \hline
   $\chi^2/{\rm dof} (a=0)$ & 7.4 & 20.6 &35.6\\ \hline
  $\chi^2/{\rm dof} (a\neq 0)$ & 0.91&  0.48 &0.58  \\ \hline
 \end{tabular}
 \end{center}
 \caption{\label{tabcont}
 Fitting values of the parameter $\hat{z}_t$ with implied values of $\Sigma$ and values of $\chi^2/{\rm dof}$ from fitting with continuum formulae of Wilson $\chi$PT. In all the quoted values, the first error is statistical, while the second one is systematic originating from the comparison of different bin sizes (taken as half of the largest difference between results using 4 different bin sizes). The range of the fit is identical to the positive $\lambda_5$-range shown in the upper, mid and lower panel of Figure \ref{figcont}. For comparison, we also give values of $\chi^2/{\rm dof}$ from Wtm$\chi$PT fits, see Figure \ref{fig1} and Table \ref{tabfit}.}
\end{table}

\begin{figure}[p!]
\begin{center}
\hspace*{-1.3cm}
\includegraphics[scale=0.272, angle =-90]{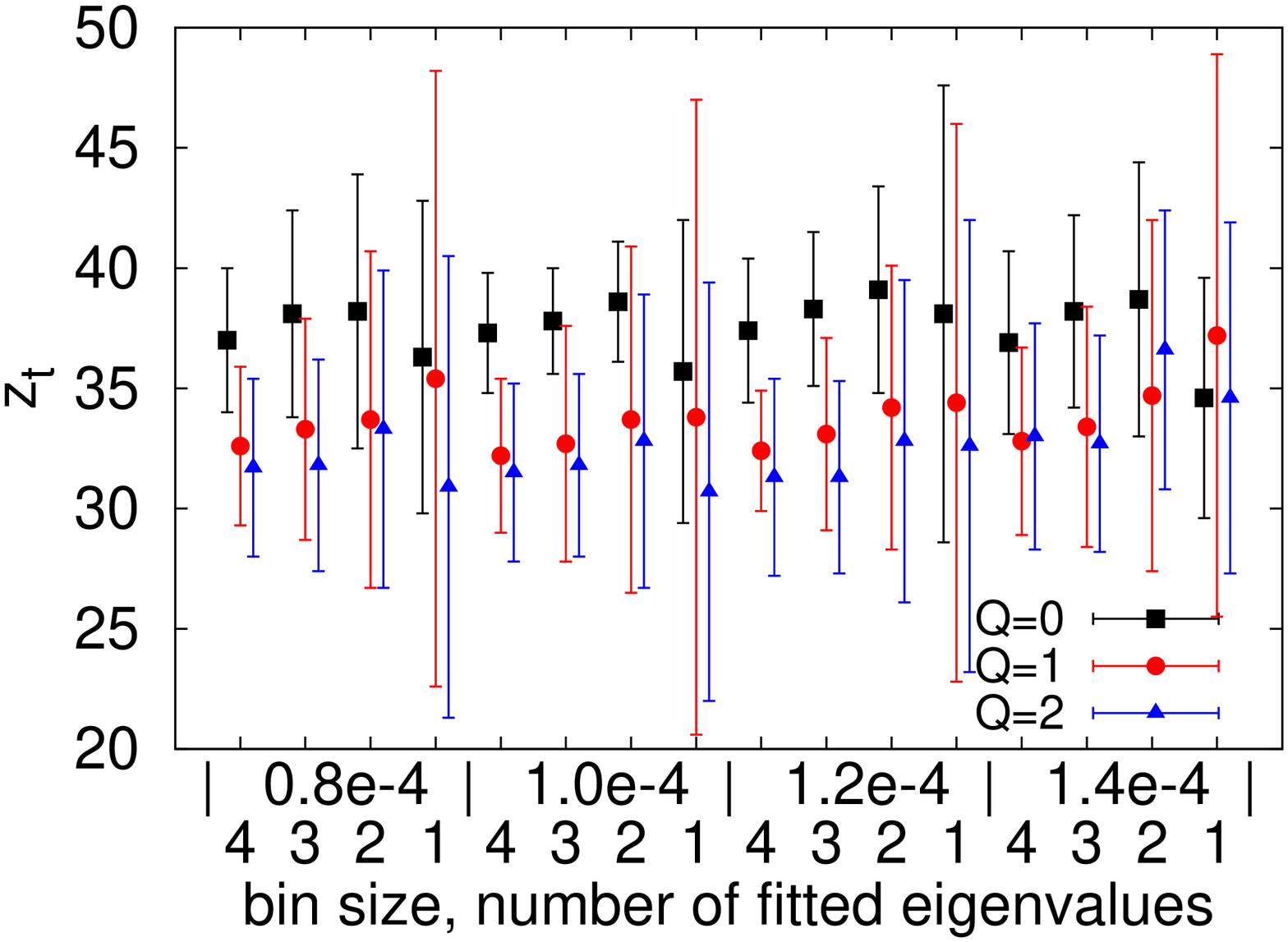}
\hspace*{-0.75cm}
\includegraphics[scale=0.272, angle =-90]{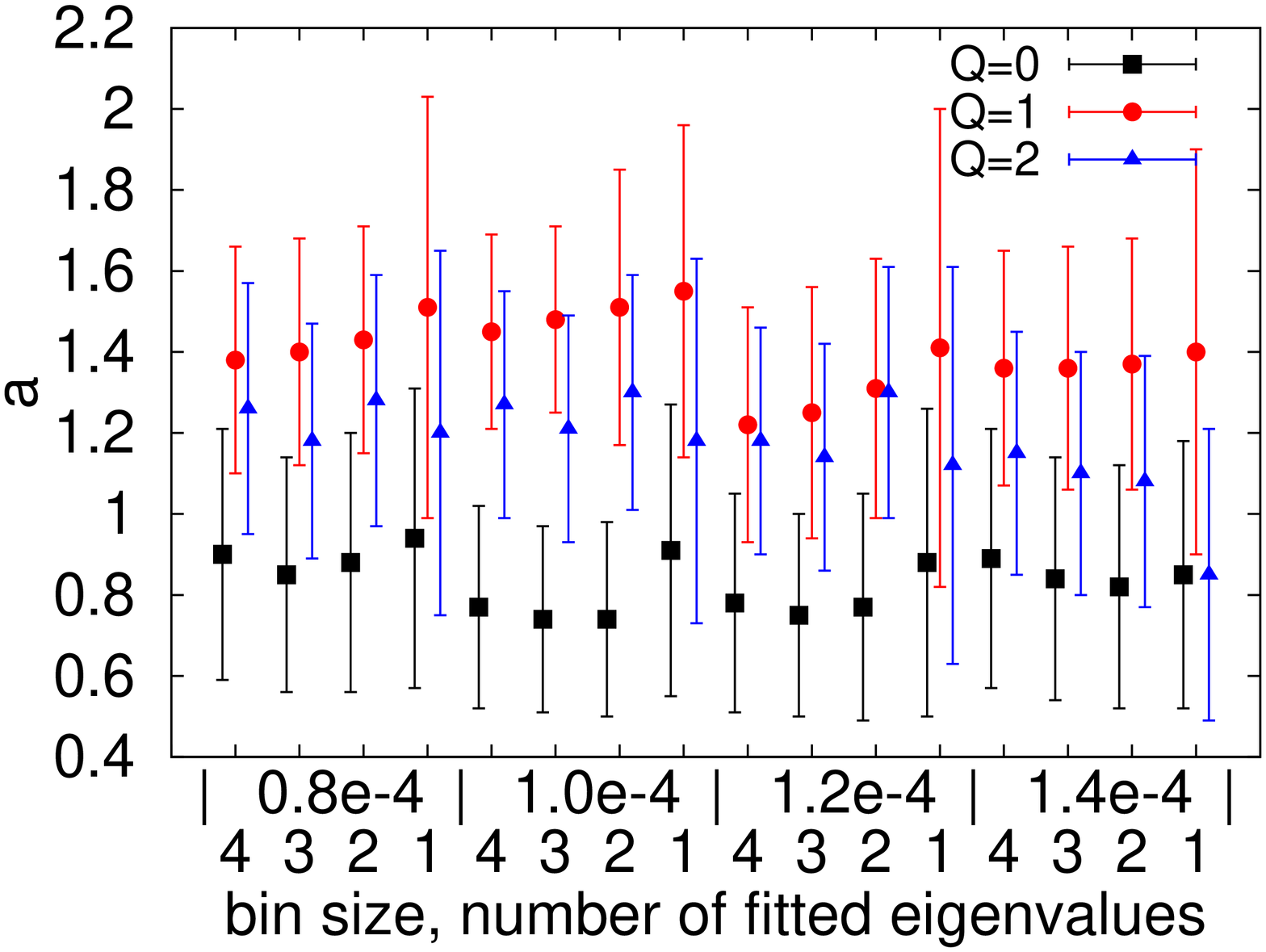}
\hspace*{-1.3cm}
\includegraphics[scale=0.272, angle =-90]{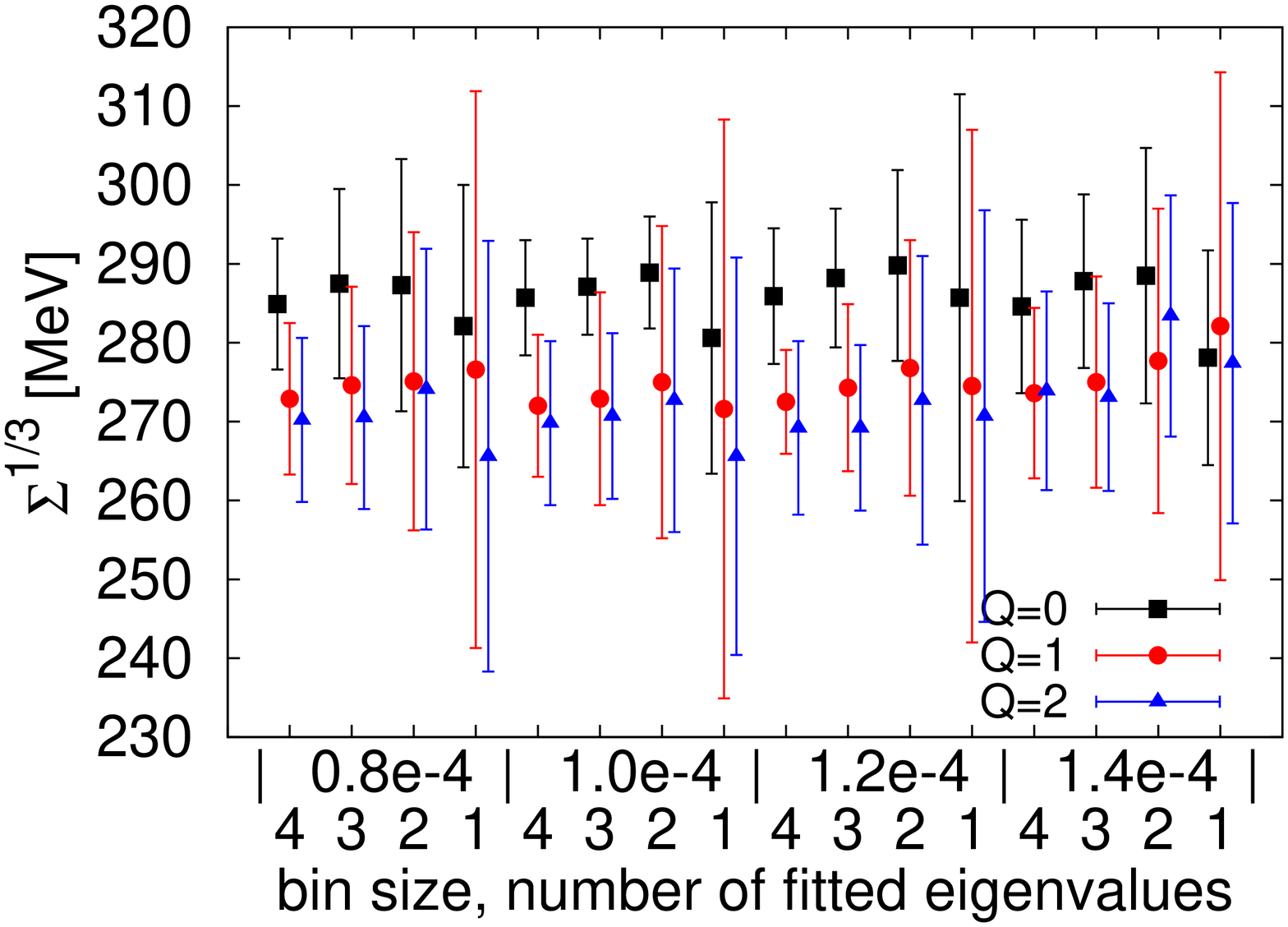}
\hspace*{-0.75cm}
\includegraphics[scale=0.272, angle =-90]{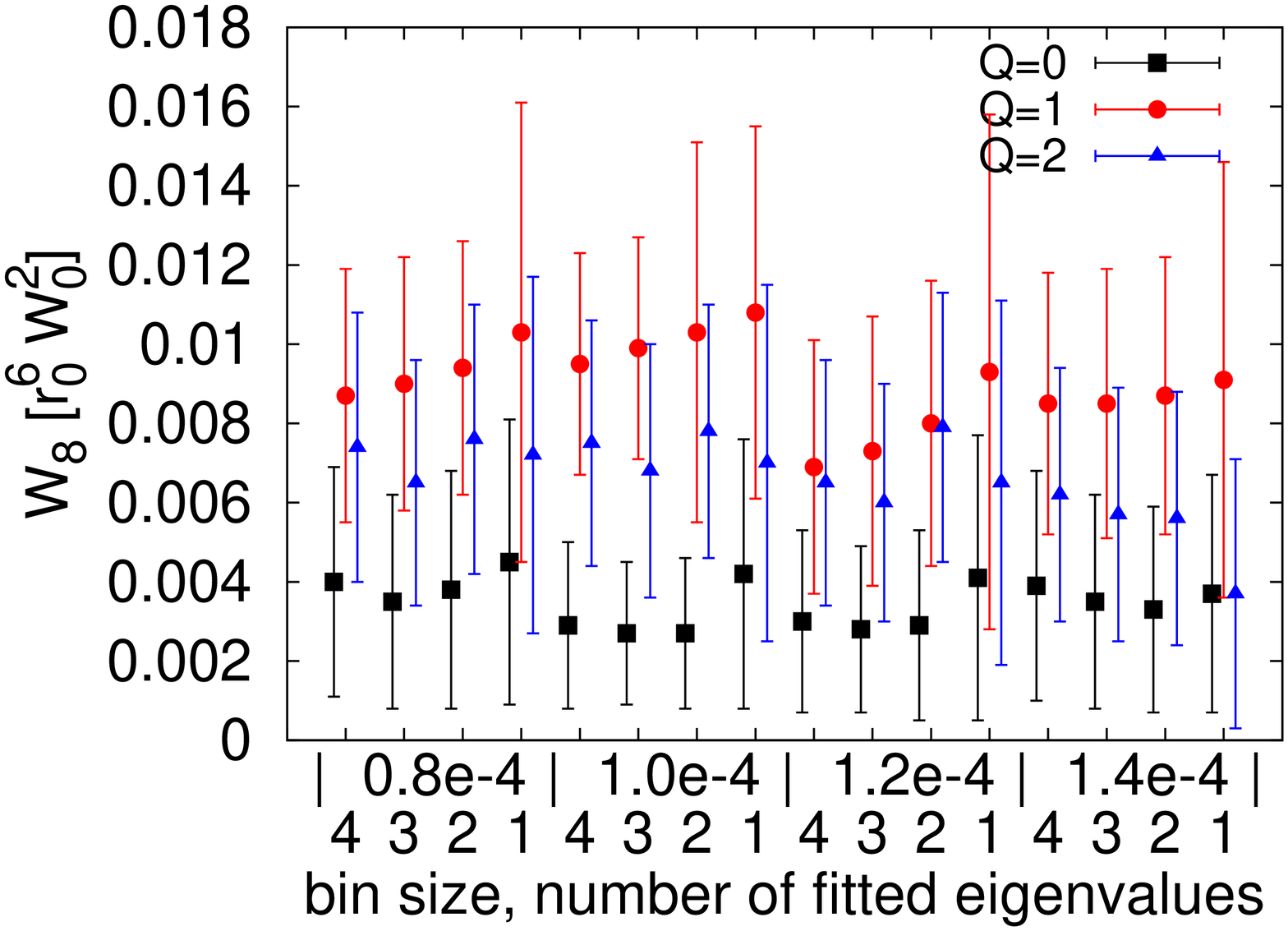}
\hspace*{-1.3cm}
\includegraphics[scale=0.272, angle =-90]{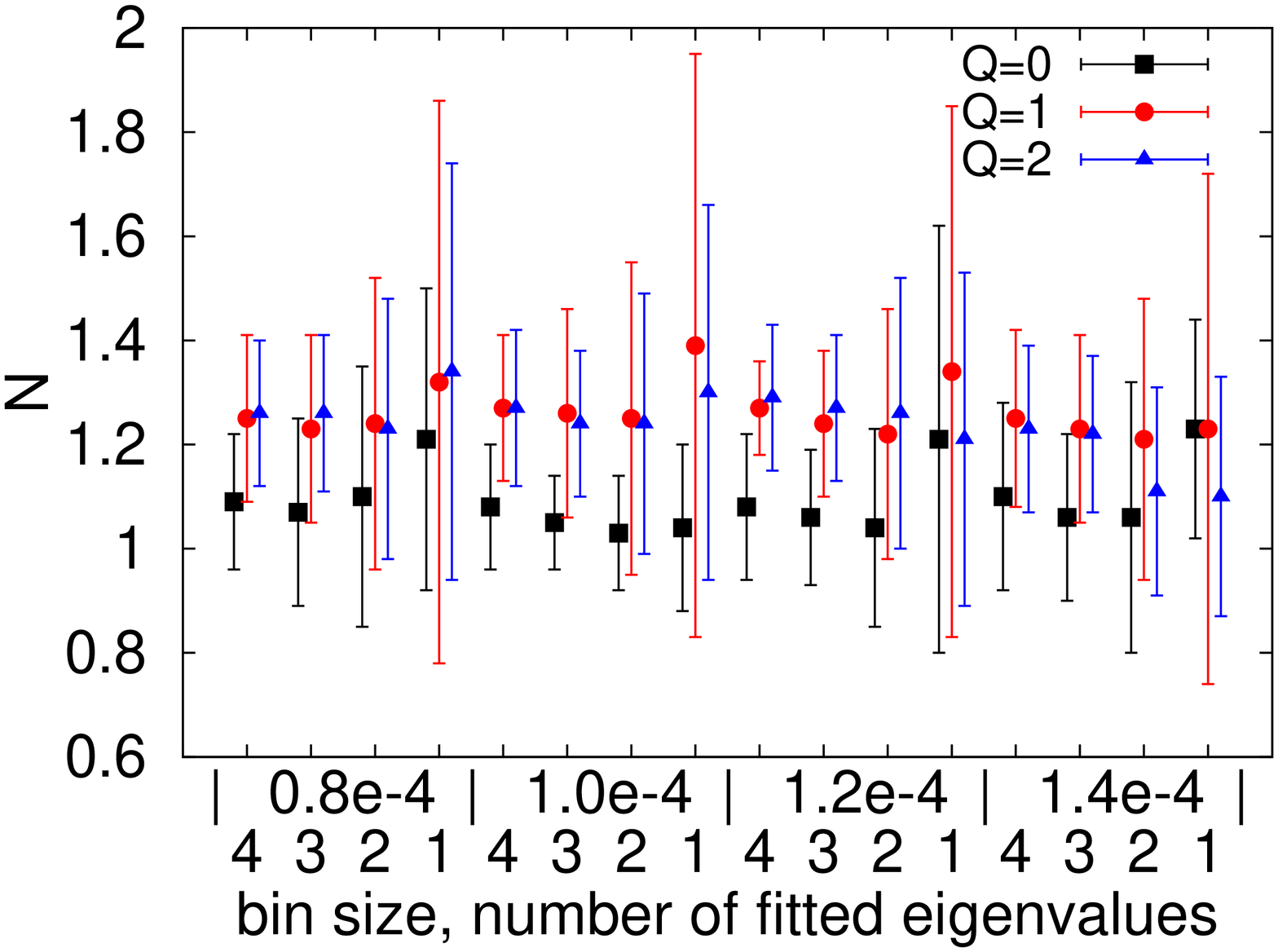}
\hspace*{-0.75cm}
\includegraphics[scale=0.272, angle =-90]{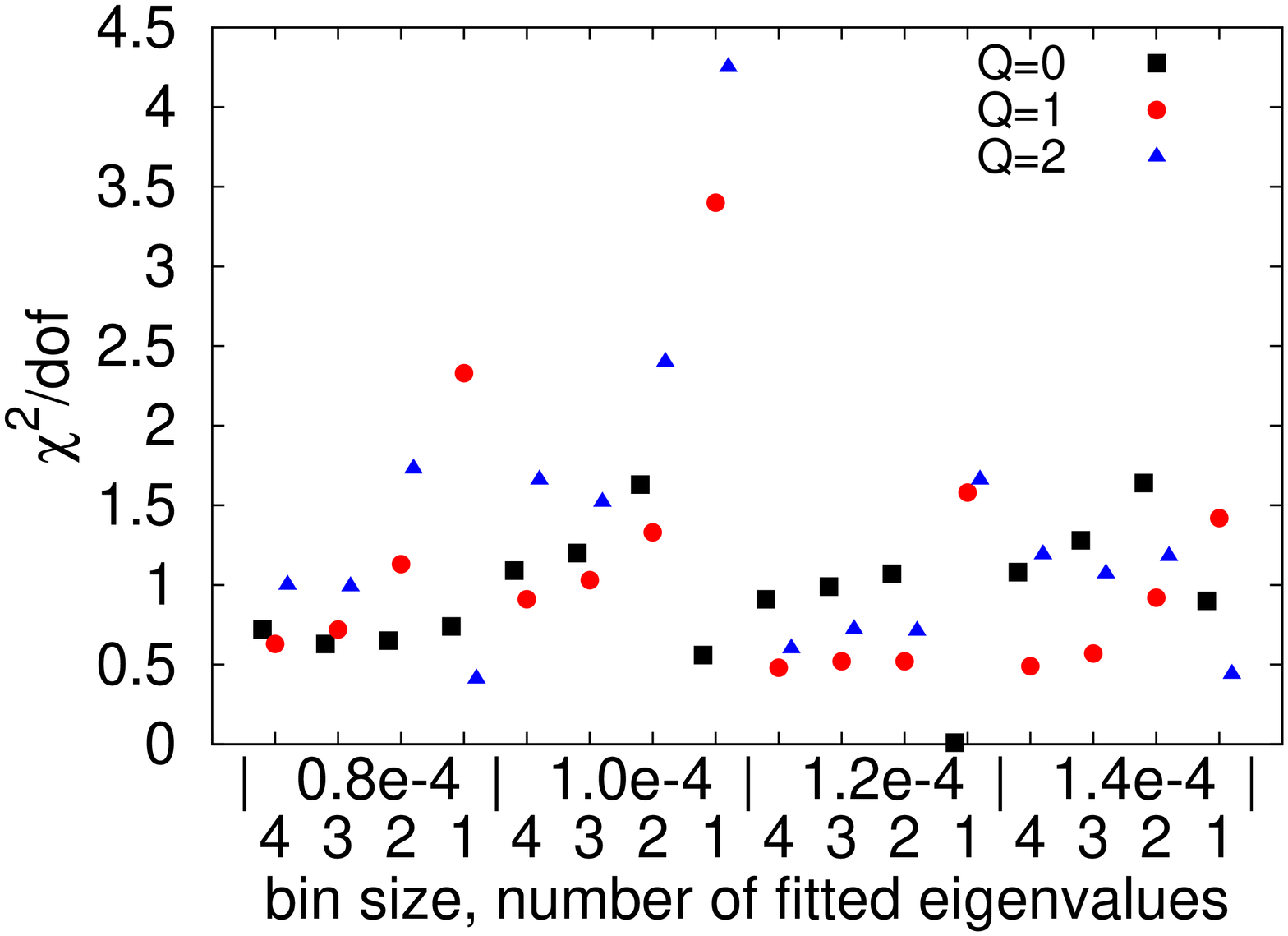}
\end{center}
\caption{\label{figparams} Fitting values of parameters $\hat{z}_t$ (top left), $\hat{a}$ (top right), $\hat{z}$ translated to $\Sigma^{1/3}$ [MeV] (middle left), $\hat{a}$ translated to $W_8$ [$r_0^6W_0^2$] (middle right), normalization $N$ (bottom left) and $\chi^2/{\rm dof}$ of the fits (bottom right). We show all analyzed variants, i.e.\ four values of the bin size, $\delta=0.8\cdot10^{-4},\,1.0\cdot10^{-4},\,1.2\cdot10^{-4},\,1.4\cdot10^{-4}$ and four fitting intervals, corresponding to considering 1, 2, 3 or 4 lowest eigenvalues of $\gamma_5 D$. The plotted errors are statistical. 
}
\end{figure}

We now discuss the extraction of LECs with our preferred fits, i.e.\ $\epsilon$-regime Wtm$\chi$PT fits with leading twisted mass cut-off effects taken into account.
As mentioned above, we use four bin sizes for our histograms and moreover, we consider four fitting ranges in $\hl^5$, corresponding to 1, 2, 3 or 4 lowest eigenvalues composing the accumulated spectral density.
Thus, in total we have 16 different fits for each of the topological charge sectors.
The values of fitting parameters for these 16 fits are shown in Figure \ref{figparams}.
We observe that the values are rather stable and the systematic differences between different fits are usually considerably lower than the statistical uncertainties.
Furthermore, we do not see any clear tendencies in the values of fitting parameters when the bin size or the fitting range are varied, apart from the tendency to larger statistical errors when the fitting range is narrowed.
As our central values, we take the ones for bin size $1.2\cdot10^{-4}$ and all four lowest eigenvalues, as the most accurate determination.
This case is illustrated in Figure \ref{fig1} and the fitting parameter values are given in Table \ref{tabfit}.
To take into account the variations introduced by different bin sizes and fitting ranges, we assign two systematic errors to each central value, apart from the statistical uncertainty.
We define these systematic errors conservatively as differences with respect to the minimum/maximum value obtained among the 16 considered fits.
As such, they are, in general, asymmetric.

We observe that there are some differences in the numerical values extracted from the different topological sectors. 
The renormalized values of $\Sigma$ that appear in Table \ref{tabfit} are in a good agreement with the results computed by ETMC with the method of spectral projectors in Ref.~\cite{KrzSigma}, where $\Sigma$ in the continuum limit was found to be, translated to physical units, $\Sigma^{1/3}\approx 290\pm 11$ MeV. 
The good agreement is very encouraging and indicates that discretization errors are taken into account to a certain degree by matching lattice results to LO Wtm$\chi$PT which only includes the single trace term, i.e.\ the one proportional to $W_8$.
At this point, it is important to say that the only real way to check this is via the analytical computation of the microscopic spectral density containing the two double trace terms, since then the residual dependence would be of $\mathcal{O}(a^4)$, and to compare the change on the extracted values of $\Sigma$. 
The value that we extract for $W_8$ is in agreement with the mixed action studies~\cite{Krzmixed}, but differs by roughly a factor of 2 from the one determined in~\cite{splittings}.

\renewcommand{\arraystretch}{1.5}
\begin{table}[t!]
\begin{center}
\begin{tabular}{ | c || c | c | c | }
   \hline $|\nu |$ & 0 & 1 &2 \\ \hline\hline
   $\hat{z}_t$ &$37.4(3.0)^{+0}_{-0.5}\,^{+1.7}_{-0}$ & $32.4(2.5)^{+0.4}_{-0.2}\,^{+2.0}_{-0}$ & $31.3(4.1)^{+1.7}_{-0}\,^{+1.5}_{-0}$\\ \hline
   $\hat{a}$ & $0.78(27)^{+12}_{-1}\,^{+10}_{-3}$ & $1.22(29)^{+23}_{-0}\,^{+19}_{-0}$ & $1.18(28)^{+9}_{-3}\,^{+12}_{-6}$ \\ \hline
   $\Sigma^{1/3}$\,\,[MeV] & $286(9)^{+0}_{-1}\,^{+4}_{-0}$ &  $273(7)^{+1}_{-1}\,^{+4}_{-0}$ & $269(11)^{+5}_{-0}\,^{+4}_{-0}$ \\ \hline
   $W_8$ \,[$r_0^6W_0^2$] & $0.0030(23)^{+10}_{-1}\,^{+11}_{-2}$&  $0.0069(32)^{+26}_{-0}\,^{+24}_{-0}$& $0.0065(31)^{+10}_{-3}\,^{+14}_{-5}$ \\ \hline
   $N$ & $1.08(14)^{+2}_{-0}\,^{+13}_{-4}$ & $1.27(9)^{+0}_{-2}\,^{+7}_{-5}$ & $1.29(14)^{+0}_{-6}\,^{+0}_{-8}$\\ \hline
   $\chi^2/{\rm dof}$ & 0.91&  0.48 &0.58  \\ \hline
 \end{tabular}
 \end{center}
 \caption{\label{tabfit}
Fitting values of the parameters $\hat{z}_t$, $\hat{a}$ and the normalization $N$, together with the extracted values of $\Sigma^{1/3}$ and $W_8$. In all the quoted values, the first error is statistical, while the asymmetric systematic ones originate from the comparison of four different bin sizes (second error) and from comparison of fits to the lowest 1, 2, 3 and 4 eigenvalues (third error). The quoted central values and the $\chi^2/{\rm dof}$ values correspond to the fits to 4 eigenvalues with bin size $1.2\cdot10^{-4}$, shown in Figure \ref{fig1}}.
\end{table}

As another check, we compared fits with and without the normalization $N$ (multiplying the eigenvalues $\lambda^5$ entering the fitting ansatz). For the case of $\nu=0$, with an extra multiplicative normalization (fitted to be $1.08(14)^{+2}_{-0}\,^{+13}_{-4}$), we obtained values shown in Table \ref{tabfit}. 
For the case of no multiplicative normalization, we obtained $z_t=39.2(9)$ and $\hat{a}=0.75(24)$, which yields $\Sigma=290.5(2.3)$ MeV and  $W_8= 0.0027(19)$ (only statistical errors), i.e.\ results compatible with the ones without the normalization.
For non-trivial topological sectors, fits without the normalization fail to describe the data completely.
The values of the normalization constant for $|\nu|=1,2$ were obtained as: $1.27(9)^{+0}_{-2}\,^{+7}_{-5}$ and $1.29(14)^{+0}_{-6}\,^{+0}_{-8}$, respectively.
This further motivates (apart from the theoretical arguments given above) the derivation of analytical formulae including $W_6$ and $W_7$.
It is important to point out that one could naively think that the low energy constant $W_7$ drops out for the case of $N_{\rm f}=2$ due to the properties of $SU(2)$ matrices ($(\tr U)^2=\tr U^2+2$) and thus, one needs to care only about the LEC $W_6$ which can actually be combined with $W_8$ to one LEC $c_2=W_6+W_8/2$. However, despite the fact that this is true for the partition function itself, it is not true for the spectral density, which is actually computed via the supersymmetric $\mathcal{Z}_{3/1}$ generating function. 

\subsection{The chGUE-GUE transition}
\begin{figure}[t!]
\begin{center}
\vspace*{-0.35cm}
\includegraphics[height=10cm, angle =-90]{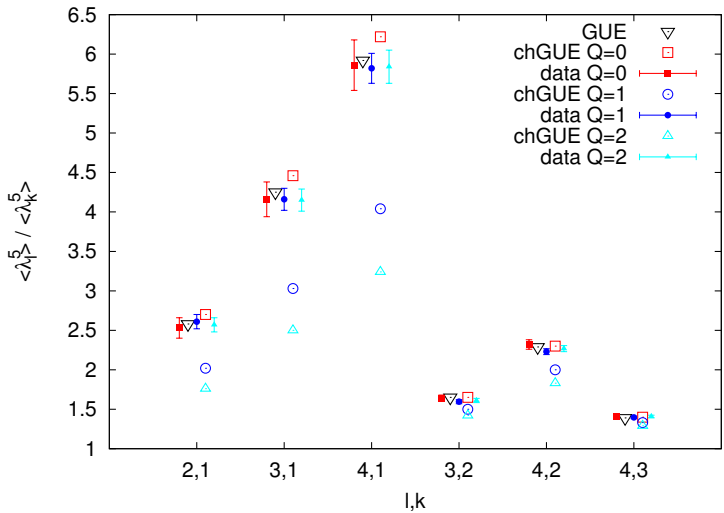}
\end{center}
\vspace*{-0.65cm}
\caption{\label{ratios} Comparison of lattice data for $\langle \lambda_l\rangle_\nu/\langle \lambda_k\rangle_\nu$ with the corresponding ratios of chGUE and symmetrized GUE. We consider the following six ratios: 2/1, 3/1, 4/1, 3/2, 4/2, 4/3. 
}
\end{figure}

The RMT model considered in this analysis can be thought of as a superposition of a chGUE (chiral Gaussian Unitary Ensemble) and of a symmetrized GUE (Gaussian Unitary Ensemble). This is easy to understand, because the lattice spacing controls the interpolation between a chiral anti-Hermitian random matrix (in the continuum) and a Hermitian matrix without any chirality properties when the lattice spacing grows large. We refer the reader to Ref.~\cite{Akemann:2011kj} for more details on this transition. In order to understand better where our parameter values stand in the given interpolation, we perform scale independent tests by considering ratios of average eigenvalues $\langle \lambda_l\rangle_\nu/\langle \lambda_k\rangle_\nu$ of our data, as in~\cite{Giusti:2003gf}. These ratios are very advantageous, since the unknown scale gets canceled in the ratio and because there is a significant cancellation of statistical noise, especially if one considers the ratio of the means of the corresponding eigenvalues. For our benchmarks, we have reproduced all the results of Table 3 of~\cite{Giusti:2003gf} by diagonalizing numerically 30000 random matrices with $n=200$ from the chGUE ensemble with $\nu=0, 1, 2$ and similarly we diagonalized  30000 $n=500$ matrices from the symmetrized GUE ensemble (the symmetrization for the case of the GUE is crucial in order to have a well defined smallest eigenvalue). In order to be able to perform a more transparent comparison of where do our results "sit" with respect to the aforementioned ensembles, we decided to show our results graphically in Figure~\ref{ratios}. As one can see, our results show in almost all cases agreement with the symmetrized GUE ensemble, which is a natural expectation, since the rescaled lattice spacing is relatively large $\mathcal{O}(1)$. Transitions between the chGUE and GUE  have a variety of exciting applications and we refer the interested reader to~\cite{Kanazawa:2018okt, Kanazawa:2018kbo} for some new results. 

\section{Conclusions}
In this article, we made the first attempt to compute the microscopic spectral density of twisted mass fermions via a direct lattice simulation. The goal was to test analytical predictions of Wilson chiral perturbation theory for twisted mass fermions and to check the feasibility of the extraction of the chiral condensate $\Sigma$ and the $W_8$, which are two LECs of the chiral Lagrangian. This study was performed within the approximation that the other two LECs $W_6$ and $W_7$, which also appear at leading order in $a^2$, are set to zero. We plan to check the validity of this assumption in an upcoming publication after we first extend the analytical result of \cite{TMSV} to include the effects described by the double trace terms (the ones involving $W_6$ and $W_7$).

Before extracting the LECs from the twisted mass $\epsilon$-regime $\chi$PT, we discussed whether the lowest eigenvalues are indeed in the $\epsilon$-regime.
We showed that there are some characteristic features of this regime that are observed in the data, in particular the considerably small overlap between single eigenvalue histograms.
Even though there are indications that we are entering the intermediate regime, we checked that excluding one eigenvalue from the fits leads to a smaller change of extracted LECs than our uncertainties -- thus, the effects of entering the intermediate regime are likely to be small and not visible at the level of LO formulae.
Moreover, we attempted fits of continuum $\epsilon$-regime formulae, concluding that they do not describe the lattice data at all.
In the end, the lattice $\epsilon$-regime formulae of Wtm$\chi$PT provide rather robust fits to the data.

The extraction of the chiral condensate $\Sigma$ is mainly controlled by the ``height" of the microscopic eigenvalue density, which we could determine fairly unambiguously and our findings are in reasonable agreement with the existing literature. However, the extraction of $W_8$ has proven to be an ordeal for almost all lattice approaches. In our study, this was quite a difficult task mainly due to the fact that for large values of $\hat{a}$, the microscopic spectral density becomes less dependent on the value of $\hat{a}$. Additionally, since the pertinent RMT for Wilson fermions (and improvements thereof) essentially describes a transition between a chGUE (for very fine lattice spacings) and a GUE (for coarser values of the lattice spacing), we studied ratios of average eigenvalues, which showed us that for the parameter values of our simulation, our results are closer to the case of the symmetrized GUE. It would still be quite challenging in the near future, even by employing the finest state-of-the art lattice ensembles, to be able to discern all the interesting features of the eigenvalue density of dynamical Wilson fermions. However, when that will be numericaly feasible, the results would be very rewarding.
A similar effect has been analytically shown for the microscopic eigenvalue density of the unimproved Wilson Dirac operator, cf.\ Figure 5 of Ref.~\cite{KVZprd}.
For this aspect, simulations at a finer lattice spacing would be very helpful.
Nevertheless, the value of $W_8$ that we extracted is in good agreement with an earlier determination from a mixed action setup, using a very different approach.

Summarizing, we believe that our work is an important first step in using dynamical twisted mass simulations to determine LECs of $\epsilon$-regime Wtm$\chi$PT.
At this stage, we were naturally limited by the availability of twisted mass configurations.
Despite using a very long ensemble, our statistical errors are rather large, thus hindering us from seeing clear evidence of $\epsilon$-regime behaviour.
In the future, the methodology developed in this paper can be used for a more clear analysis, using significantly more statistics and/or using lattice parameters where more sensitivity is expected in the analytical formulae, i.e.\ at a finer lattice spacing (still with a sufficiently large volume) and with a smaller pion mass.
On the theoretical side, more robustness is also expected when possibly non-zero values of the LECs $W_6$ and $W_7$ are taken into consideration.

\section{Acknowledgements}

This study was based on a variant of the ETMC's public lattice Quantum Chromodynamics  code, tmLQCD \cite{tmlqcd1, tmlqcd2}.
We would like to thank Poul Damgaard, Elena Garcia-Ramos, Gregorio Herdoiza, Karl Jansen, Mario Kieburg, Joyce C. Myers, Steve Sharpe, Andrea Shindler and in particular Jac Verbaarschot for fruitful discussions, Kim Splittorff for collaboration during early stages of this work and Andreas Athenodorou for providing us with the data of the topological charge.
This work was granted access to the HPC resources of CINES and IDRIS under the allocations offered by GENCI. We express our gratitude to the staff of this Computing facility for their constant help.
This work was supported by the DFG Collaborative Research Centre SFB 1225
(ISOQUANT), the Humboldt Foundation, the National Science Foundation (USA) under grant PHY-1516509, by the Jefferson Science Associates, LLC under U.S. DOE Contract \#DE-AC05-06OR23177 (S.Z.). K.C.\ was supported in part by the Deutsche Forschungsgemeinschaft (DFG), project nr. CI 236/1-1. 




\begin{thebibliography}{99}
\bibitem{FrezzottiA} 
  R.~Frezzotti {\it et al.} [Alpha Collaboration],
  ``Lattice QCD with a chirally twisted mass term,''
  JHEP {\bf 0108}, 058 (2001)
  [hep-lat/0101001].


\bibitem{Frezzossi} 
  R.~Frezzotti and G.~C.~Rossi,
  ``Chirally improving Wilson fermions. 1. O(a) improvement,''
  JHEP {\bf 0408}, 007 (2004)
  [hep-lat/0306014].
  
\bibitem{Frezzossi2} 
  R.~Frezzotti and G.~C.~Rossi,
  ``Chirally improving Wilson fermions. II. Four-quark operators,''
  JHEP {\bf 0410}, 070 (2004)
  [hep-lat/0407002].
  
\bibitem{Sint} 
  S.~Sint,
  ``Lattice QCD with a chiral twist,''
  hep-lat/0702008.
  
  
\bibitem{Andrea} 
  A.~Shindler,
  ``Twisted mass lattice QCD,''
  Phys.\ Rept.\  {\bf 461}, 37 (2008)
  [arXiv:0707.4093 [hep-lat]].
  
\bibitem{PP1} 
  A.~Abdel-Rehim {\it et al.},
  ``Nucleon and pion structure with lattice QCD simulations at physical value of the pion mass,''
  Phys.\ Rev.\ D {\bf 92}, no. 11, 114513 (2015)
  Erratum: [Phys.\ Rev.\ D {\bf 93}, no. 3, 039904 (2016)]
  [arXiv:1507.04936 [hep-lat]].
  
\bibitem{PP2} 
  A.~Abdel-Rehim {\it et al.} [ETM Collaboration],
  ``Simulating QCD at the Physical Point with $N_{\rm f}=2$ Wilson Twisted Mass Fermions at Maximal Twist,''
  arXiv:1507.05068 [hep-lat].

\bibitem{slowing} 
  S.~Schaefer {\it et al.} [ALPHA Collaboration],
  ``Critical slowing down and error analysis in lattice QCD simulations,''
  Nucl.\ Phys.\ B {\bf 845}, 93 (2011)
  [arXiv:1009.5228 [hep-lat]].

\bibitem{LSopen} 
  M.~Luscher and S.~Schaefer,
  ``Lattice QCD without topology barriers,''
  JHEP {\bf 1107}, 036 (2011)
  [arXiv:1105.4749 [hep-lat]].
  
  
\bibitem{SS} 
  S.~R.~Sharpe and R.~L.~Singleton, Jr,
  ``Spontaneous flavor and parity breaking with Wilson fermions,''
  Phys.\ Rev.\ D {\bf 58}, 074501 (1998)
  [hep-lat/9804028].
      
  
  
\bibitem{RS} 
  G.~Rupak and N.~Shoresh,
  ``Chiral perturbation theory for the Wilson lattice action,''
  Phys.\ Rev.\ D {\bf 66}, 054503 (2002)
  [hep-lat/0201019].

\bibitem{BRS} 
  O.~Bar, G.~Rupak and N.~Shoresh,
  ``Chiral perturbation theory at O(a**2) for lattice QCD,''
  Phys.\ Rev.\ D {\bf 70}, 034508 (2004)
  [hep-lat/0306021].
  
\bibitem{SharpeNara} 
  S.~R.~Sharpe,
  ``Applications of Chiral Perturbation theory to lattice QCD,''
  hep-lat/0607016.
  
\bibitem{SharpeWu} 
  S.~R.~Sharpe and J.~M.~S.~Wu,
  ``The Phase diagram of twisted mass lattice QCD,''
  Phys.\ Rev.\ D {\bf 70}, 094029 (2004)
  [hep-lat/0407025].
  
  S.~R.~Sharpe and J.~M.~S.~Wu,
  ``Applying chiral perturbation to twisted mass lattice QCD,''
  Nucl.\ Phys.\ Proc.\ Suppl.\  {\bf 140}, 323 (2005)
  [hep-lat/0407035].
  
  
\bibitem{Scorzato} 
  L.~Scorzato,
  ``Pion mass splitting and phase structure in twisted mass QCD,''
  Eur.\ Phys.\ J.\ C {\bf 37}, 445 (2004)
  [hep-lat/0407023].
  
  
\bibitem{Muenster} 
  G.~Munster,
  ``On the phase structure of twisted mass lattice QCD,''
  JHEP {\bf 0409}, 035 (2004)
  [hep-lat/0407006].
  
  
  
\bibitem{KSVZ} 
  M.~Kieburg, K.~Splittorff, J.~J.~M.~Verbaarschot and S.~Zafeiropoulos,
  ``Phase Diagram of Wilson and Twisted Mass Fermions at finite isospin chemical potential,''
  PoS LATTICE {\bf 2014}, 065 (2015)
  [arXiv:1411.2570 [hep-lat]].
  
\bibitem{JKSVZ} 
  O.~Janssen, M.~Kieburg, K.~Splittorff, J.~J.~M.~Verbaarschot and S.~Zafeiropoulos,
  ``Phase Diagram of Dynamical Twisted Mass Wilson Fermions at Finite Isospin Chemical Potential,''
  Phys.\ Rev.\ D {\bf 93} (2016) no.9,  094502
  [arXiv:1509.02760 [hep-lat]].
  
\bibitem{Aoki} 
  S.~Aoki,
  ``New Phase Structure for Lattice QCD with Wilson Fermions,''
  Phys.\ Rev.\ D {\bf 30}, 2653 (1984).
  
\bibitem{realization} 
  M.~Kieburg, K.~Splittorff and J.~J.~M.~Verbaarschot,
  ``The Realization of the Sharpe-Singleton Scenario,''
  Phys.\ Rev.\ D {\bf 85}, 094011 (2012)
  [arXiv:1202.0620 [hep-lat]].
  

\bibitem{DSV} 
  P.~H.~Damgaard, K.~Splittorff and J.~J.~M.~Verbaarschot,
  ``Microscopic Spectrum of the Wilson Dirac Operator,''
  Phys.\ Rev.\ Lett.\  {\bf 105}, 162002 (2010)
  [arXiv:1001.2937 [hep-th]].
  
  
  
  
  
\bibitem{ADSV} 
  G.~Akemann, P.~H.~Damgaard, K.~Splittorff and J.~J.~M.~Verbaarschot,
  ``Spectrum of the Wilson Dirac Operator at Finite Lattice Spacings,''
  Phys.\ Rev.\ D {\bf 83}, 085014 (2011)
  [arXiv:1012.0752 [hep-lat]].
  
  
  


\bibitem{KVZprl} 
  M.~Kieburg, J.~J.~M.~Verbaarschot and S.~Zafeiropoulos,
  ``Eigenvalue Density of the non-Hermitian Wilson Dirac Operator,''
  Phys.\ Rev.\ Lett.\  {\bf 108}, 022001 (2012)
  [arXiv:1109.0656 [hep-lat]].
  
\bibitem{Hansen} 
  M.~T.~Hansen and S.~R.~Sharpe,
  ``Constraint on the Low Energy Constants of Wilson Chiral Perturbation Theory,''
  Phys.\ Rev.\ D {\bf 85}, 014503 (2012)
  [arXiv:1111.2404 [hep-lat]].
  
\bibitem{KVZprd} 
  M.~Kieburg, J.~J.~M.~Verbaarschot and S.~Zafeiropoulos,
  ``Spectral Properties of the Wilson Dirac Operator and random matrix theory,''
  Phys.\ Rev.\ D {\bf 88}, 094502 (2013)
  [arXiv:1307.7251 [hep-lat]].
\bibitem{KVZ2c} 
  M.~Kieburg, J.~J.~M.~Verbaarschot and S.~Zafeiropoulos,
  ``Dirac Spectrum of the Wilson Dirac Operator for QCD with Two Colors,''
  Phys.\ Rev.\ D {\bf 92}, no. 4, 045026 (2015)
  [arXiv:1505.01784 [hep-lat]].
  
\bibitem{AokiBar} 
  S.~Aoki, O.~Bar and B.~Biedermann,
  ``Pion scattering in Wilson chiral perturbation theory,''
  Phys.\ Rev.\ D {\bf 78}, 114501 (2008)
  [arXiv:0806.4863 [hep-lat]].
  
\bibitem{B1} 
  O.~Bar, S.~Necco and A.~Shindler,
  ``The epsilon regime with twisted mass Wilson fermions,''
  JHEP {\bf 1004}, 053 (2010)
  [arXiv:1002.1582 [hep-lat]].
  
\bibitem{B2} 
  O.~Bar, S.~Necco and S.~Schaefer,
  ``The Epsilon regime with Wilson fermions,''
  JHEP {\bf 0903}, 006 (2009)
  [arXiv:0812.2403 [hep-lat]].
  
\bibitem{B3} 
  O.~B\"ar and B.~H\"orz,
  ``Charmless chiral perturbation theory for $N_{\rm f}=2+1+1$ twisted mass lattice QCD,''
  Phys.\ Rev.\ D {\bf 90}, no. 3, 034508 (2014)
  [arXiv:1402.6145 [hep-lat]].
  
\bibitem{Krzmixed} 
  K.~Cichy, V.~Drach, E.~Garcia-Ramos, G.~Herdoiza and K.~Jansen,
  ``Overlap valence quarks on a twisted mass sea: a case study for mixed action Lattice QCD,''
  Nucl.\ Phys.\ B {\bf 869}, 131 (2013)
  [arXiv:1211.1605 [hep-lat]].
  
\bibitem{splittings} 
  G.~Herdoiza, K.~Jansen, C.~Michael, K.~Ottnad and C.~Urbach,
  ``Determination of Low-Energy Constants of Wilson Chiral Perturbation Theory,''
  JHEP {\bf 1305}, 038 (2013)
  [arXiv:1303.3516 [hep-lat]].
  
\bibitem{Fabio} 
  F.~Bernardoni, J.~Bulava and R.~Sommer,
  ``Determination of the Wilson ChPT low energy constant $c_2$,''
  PoS LATTICE {\bf 2011}, 095 (2011)
  [arXiv:1111.4351 [hep-lat]].
  
\bibitem{DHS1} 
  P.~H.~Damgaard, U.~M.~Heller and K.~Splittorff,
  ``Finite-Volume Scaling of the Wilson Dirac Operator Spectrum,''
  Phys.\ Rev.\ D {\bf 85}, 014505 (2012)
  [arXiv:1110.2851 [hep-lat]].
  
  
\bibitem{DWW} 
  A.~Deuzeman, U.~Wenger and J.~Wuilloud,
  ``Spectral properties of the Wilson Dirac operator in the $\epsilon$-regime,''
  JHEP {\bf 1112}, 109 (2011)
  [arXiv:1110.4002 [hep-lat]].
  
  
\bibitem{DHS2} 
  P.~H.~Damgaard, U.~M.~Heller and K.~Splittorff,
  ``New Ways to Determine Low-Energy Constants with Wilson Fermions,''
  Phys.\ Rev.\ D {\bf 86}, 094502 (2012)
  [arXiv:1206.4786 [hep-lat]].
  
\bibitem{TMSV} 
  K.~Splittorff and J.~J.~M.~Verbaarschot,
  ``The Microscopic Twisted Mass Dirac Spectrum,''
  Phys.\ Rev.\ D {\bf 85}, 105008 (2012)
  [arXiv:1201.1361 [hep-lat]].

 
\bibitem{lat15proc} 
   K.~Cichy, E.~Garcia-Ramos, K.~Splittorff and S.~Zafeiropoulos,
  ``The microscopic Twisted Mass Dirac spectrum and the spectral determination of the LECs of Wilson $\chi$PT,''
  PoS LATTICE {\bf 2015} (2016) 058
  [arXiv:1510.09169 [hep-lat]].
  
\bibitem{proc2} 
  K.~Cichy, E.~Garcia-Ramos, K.~Splittorff and S.~Zafeiropoulos,
  ``Twisted Mass Wilson $\chi $-PT Versus Lattice Data: a Case Study,''
  Acta Phys.\ Polon.\ Supp.\  {\bf 9}, 427 (2016).

\bibitem{Shindler:2009ri}
  A.~Shindler,
  ``Observations on the Wilson fermions in the epsilon regime,''
  Phys.\ Lett.\ B {\bf 672} (2009) 82
  [arXiv:0812.2251 [hep-lat]].  
  
\bibitem{GS} 
  C.~Gattringer and S.~Solbrig,
  ``Remnant index theorem and low-lying eigenmodes for twisted mass fermions,''
  Phys.\ Lett.\ B {\bf 621}, 195 (2005)
  [hep-lat/0503004].
  

\bibitem{SW} 
  B.~Sheikholeslami and R.~Wohlert,
  ``Improved Continuum Limit Lattice Action for QCD with Wilson Fermions,''
  Nucl.\ Phys.\ B {\bf 259}, 572 (1985).
\bibitem{SVdyn} 
  K.~Splittorff and J.~J.~M.~Verbaarschot,
  ``The Wilson Dirac Spectrum for QCD with Dynamical Quarks,''
  Phys.\ Rev.\ D {\bf 84}, 065031 (2011)
  [arXiv:1105.6229 [hep-lat]].
\bibitem{SmitVink} 
  J.~Smit and J.~C.~Vink,
  ``Remnants of the Index Theorem on the Lattice,''
  Nucl.\ Phys.\ B {\bf 286}, 485 (1987).
  
\bibitem{Itoh} 
  S.~Itoh, Y.~Iwasaki and T.~Yoshie,
  ``The U(1) Problem and Topological Excitations on a Lattice,''
  Phys.\ Rev.\ D {\bf 36}, 527 (1987).
  
\bibitem{EHN} 
  R.~G.~Edwards, U.~M.~Heller, J.~E.~Kiskis and R.~Narayanan,
  ``Quark spectra, topology and random matrix theory,''
  Phys.\ Rev.\ Lett.\  {\bf 82}, 4188 (1999)
  [hep-th/9902117].
  
  
  
  
  

  
\bibitem{KaiserLeutwyler} 
  R.~Kaiser and H.~Leutwyler,
  ``Large N(c) in chiral perturbation theory,''
  Eur.\ Phys.\ J.\ C {\bf 17}, 623 (2000)
  [hep-ph/0007101].
    
  
 
   
\bibitem{Baron1} 
  R.~Baron {\it et al.},
  ``Light hadrons from lattice QCD with light (u,d), strange and charm dynamical quarks,''
  JHEP {\bf 1006}, 111 (2010)
  [arXiv:1004.5284 [hep-lat]].
  
\bibitem{Baron2} 
  R.~Baron {\it et al.} [ETM Collaboration],
  ``Computing K and D meson masses with $N_{f}$ = 2+1+1 twisted mass lattice QCD,''
  Comput.\ Phys.\ Commun.\  {\bf 182}, 299 (2011)
  [arXiv:1005.2042 [hep-lat]].
  
%
%
%
  
  
\bibitem{iwasaki1} 
  Y.~Iwasaki,
  ``Renormalization Group Analysis of Lattice Theories and Improved Lattice Action: Two-Dimensional Nonlinear O(N) Sigma Model,''
  Nucl.\ Phys.\ B {\bf 258}, 141 (1985).
  
  
 \bibitem{iwasaki2}
   Y.~Iwasaki, K.~Kanaya, T.~Kaneko and T.~Yoshie,
   ``Scaling in SU(3) pure gauge theory with a renormalization group improved action,''
   Phys.\ Rev.\ D {\bf 56}, 151 (1997)
   [hep-lat/9610023].

\bibitem{Frezzotti:2003xj}
  R.~Frezzotti and G.~C.~Rossi,
  ``Twisted mass lattice QCD with mass nondegenerate quarks,''
  Nucl.\ Phys.\ Proc.\ Suppl.\  {\bf 128} (2004) 193
  [hep-lat/0311008].     
     
 \bibitem{italiani}
   N.~Carrasco {\it et al.} [ETM Collaboration],
   ``Up, down, strange and charm quark masses with N$_f$ = 2+1+1 twisted mass lattice QCD,''
   Nucl.\ Phys.\ B {\bf 887}, 19 (2014)
   [arXiv:1403.4504 [hep-lat]].
  
\bibitem{jamesjac} 
  J.~C.~Osborn and J.~J.~M.~Verbaarschot,
  ``Thouless energy and correlations of QCD Dirac eigenvalues,''
  Phys.\ Rev.\ Lett.\  {\bf 81}, 268 (1998)
  [hep-ph/9807490].
  

%
%
%
%
  
\bibitem{WF} 
  M.~L\"uscher,
  ``Properties and uses of the Wilson flow in lattice QCD,''
  JHEP {\bf 1008}, 071 (2010)
  [JHEP {\bf 1403}, 092 (2014)]
  [arXiv:1006.4518 [hep-lat]].
\bibitem{local} 
  P.~Hernandez, K.~Jansen and M.~Luscher,
  Nucl.\ Phys.\ B {\bf 552}, 363 (1999)
  doi:10.1016/S0550-3213(99)00213-8
  [hep-lat/9808010].
  
  \bibitem{marga} 
  P.~de Forcrand, M.~Garcia Perez and I.~O.~Stamatescu,
  ``Topology of the SU(2) vacuum: A Lattice study using improved cooling,''
  Nucl.\ Phys.\ B {\bf 499}, 409 (1997)
  [hep-lat/9701012].

  
\bibitem{andreas} 
  C.~Alexandrou, A.~Athenodorou and K.~Jansen,
  ``Topological charge using cooling and the gradient flow,''
  Phys.\ Rev.\ D {\bf 92}, no. 12, 125014 (2015)
  [arXiv:1509.04259 [hep-lat]].

\bibitem{Alexandrou:2017hqw}
  C.~Alexandrou, A.~Athenodorou, K.~Cichy, A.~Dromard, E.~Garcia-Ramos, K.~Jansen, U.~Wenger and F.~Zimmermann,
  ``Comparison of topological charge definitions in Lattice QCD,''
  arXiv:1708.00696 [hep-lat].  
  
\bibitem{cubature}
  S.~G.~Johnson, ``cubature'' library, available at:
  \verb|https://github.com/stevengj/cubature|.
  
\bibitem{Berntsen}  
J.~Berntsen, T.~O.~Espelid and A.~Genz, 
``An adaptive algorithm for the approximate calculation of multiple integrals,''
     ACM Trans. Math. Soft. 17 (4), 437-451 (1991).  

\bibitem{cyprus} 
  C.~Alexandrou, M.~Constantinou, H.~Panagopoulos [ETM Collaboration],
  ``Renormalization functions for Nf=2 and Nf=4 Twisted Mass fermions,''
  arXiv:1509.00213 [hep-lat].
   
 \bibitem{francais}
   B.~Blossier {\it et al.} [ETM Collaboration],
   ``Renormalization of quark propagator, vertex functions, and twist-2 operators from twisted-mass lattice QCD at $N_{\rm f}$=4,''
   Phys.\ Rev.\ D {\bf 91}, no. 11, 114507 (2015)
   [arXiv:1411.1109 [hep-lat]].
     
\bibitem{Banks:1979yr}
  T.~Banks and A.~Casher,
  ``Chiral Symmetry Breaking in Confining Theories,''
  Nucl.\ Phys.\ B {\bf 169} (1980) 103.     
     
\bibitem{Necco:2011vx}
  S.~Necco and A.~Shindler,
  ``Spectral density of the Hermitean Wilson Dirac operator: a NLO computation in chiral perturbation theory,''
  JHEP {\bf 1104} (2011) 031     


\bibitem{Akemann:2012pn}
  G.~Akemann and A.~C.~Ipsen,
  ``Individual Eigenvalue Distributions for the Wilson Dirac Operator,''
  JHEP {\bf 1204} (2012) 102
  [arXiv:1202.1241 [hep-lat]].
  
  


   
 
 \bibitem{KrzSigma}
   K.~Cichy, E.~Garcia-Ramos and K.~Jansen,
   ``Chiral condensate from the twisted mass Dirac operator spectrum,''
   JHEP {\bf 1310}, 175 (2013)
   [arXiv:1303.1954 [hep-lat]].
\bibitem{KrzSigma2} 
  C.~Alexandrou, A.~Athenodorou, K.~Cichy, M.~Constantinou, D.~P.~Horkel, K.~Jansen, G.~Koutsou and C.~Larkin,
  arXiv:1709.06596 [hep-lat].

\bibitem{Akemann:2011kj} 
  G.~Akemann and T.~Nagao,
  JHEP {\bf 1110}, 060 (2011)
  doi:10.1007/JHEP10(2011)060
  [arXiv:1108.3035 [math-ph]].
  
\bibitem{Giusti:2003gf} 
  L.~Giusti, M.~Luscher, P.~Weisz and H.~Wittig,
  JHEP {\bf 0311}, 023 (2003)
  doi:10.1088/1126-6708/2003/11/023
  [hep-lat/0309189].
  
\bibitem{Verbaarschot:1994ip} 
  J.~J.~M.~Verbaarschot and I.~Zahed,
  Phys.\ Rev.\ Lett.\  {\bf 73}, 2288 (1994)
  doi:10.1103/PhysRevLett.73.2288
  [hep-th/9405005].
  
\bibitem{Kanazawa:2018okt} 
  T.~Kanazawa and M.~Kieburg,
  Phys.\ Rev.\ Lett.\  {\bf 120}, no. 24, 242001 (2018)
  doi:10.1103/PhysRevLett.120.242001
  [arXiv:1803.04122 [hep-th]].
  
  
  
\bibitem{Kanazawa:2018kbo} 
  T.~Kanazawa and M.~Kieburg,
  J.\ Phys.\ A {\bf 51}, no. 34, 345202 (2018)
  doi:10.1088/1751-8121/aace3b
  [arXiv:1804.03985 [math-ph]].
   
\bibitem{tmlqcd1}
  K.~Jansen and C.~Urbach,
  ``tmLQCD: A Program suite to simulate Wilson Twisted mass Lattice QCD,''
  Comput.\ Phys.\ Commun.\  {\bf 180}, 2717 (2009)
  [arXiv:0905.3331 [hep-lat]].

\bibitem{tmlqcd2}
  C.~Urbach, K.~Jansen, A.~Shindler and U.~Wenger,
  ``HMC algorithm with multiple time scale integration and mass preconditioning,''
  Comput.\ Phys.\ Commun.\  {\bf 174}, 87 (2006)
  [hep-lat/0506011].
  
\end{thebibliography}
\end{document}